# Alkali-silica reaction products and cracks: X-ray micro-tomography-based analysis of their spatial-temporal evolution at a mesoscale

Mahdieh Shakoorioskooie[a,b,c,♦], Michele Griffa[a,♦], Andreas Leemann[a], Robert Zboray[b], Pietro Lura[a,c]

[a] *Concrete and Asphalt Laboratory, Empa, Swiss Federal Laboratories for Materials Science and Technology, CH-8600 Dübendorf, Switzerland*

[b] *Center for X-ray Analytics, Empa, Swiss Federal Laboratories for Materials Science and Technology, CH-8600 Dübendorf, Switzerland*

[c] *Institute for Building Materials (IfB), ETH Zürich, CH-8093 Zürich, Switzerland*

**Abstract**

In this study, we propose a laboratory-scale methodology, based on X-ray micro-tomography and caesium (Cs) as a contrast agent, to advance the understanding of cracking due to alkali-silica reaction (ASR) in concrete. The methodology allows achieving a completely non-destructive and time-lapse characterization of the spatial-temporal patterns of both the cracks and the ASR products. While Cs addition slightly accelerated the ASR kinetics, the crack patterns, with and without Cs addition, were statistically equivalent. Cracks with ASR products appeared first in the aggregates, close to the interface with the cement paste. They propagated afterwards towards the aggregates interior. Some products were then extruded for several mm into air voids and cracks in the cement paste. This process suggests that, in the early stage, the ASR products may be a low-viscosity gel that can flow away from the source aggregate and may settle later elsewhere as a rigid phase, upon calcium uptake.

## 1. INTRODUCTION

Amongst various concrete durability issues, cracking induced by the alkali-silica reaction (ASR) is one of the most deleterious problems. ASR consists of reactions between metastable

---

[♦] Corresponding authors.
E-Mail addresses: mahdieh.shakoorioskooie@empa.ch; michele.griffa@empa.ch

silica (present in many aggregates) and highly alkaline concrete pore solution [1]. The formation of ASR products is associated with the development of mechanical stresses and subsequent cracking. Depending on the environmental conditions (temperature, humidity, alkali concentration and availability), ASR can lead to widespread and pervasive cracking through a concrete structure, within a time scale of years or decades, potentially jeopardizing its integrity, durability and functionality [2]. ASR damage usually appears as interconnected crack networks throughout the entire structure, which become also visible on the external surfaces [3].

Several recent studies have focused on some of the still-unknown features of ASR, such as initial product formation [4], secondary stage product distribution [5] and the corresponding cracking [6]–[8]. Among different characterization methods to study such aspects of ASR cracking, methods based on imaging have been playing a significant role. For instance, optical and electron microscopy have been widely implemented to visualize ASR products and cracks with high (i.e., micron and sub-micron) spatial resolution [9], [10]. Nevertheless, such microscopy techniques require a destructive sample preparation, thus preventing time-lapse, continuous monitoring of products and cracks evolutions for the exact same regions within a specimen. The latter is a significant hindrance to improving the understanding of different ASR features and their development. To overcome these limitations, using a non-destructive imaging technique such as X-ray micro-tomography (XMT) is crucial. The use of XMT to characterize ASR and its associated cracking has been reported in few cases and for limited types of cement-based materials, such as glass-alkali solution mixture systems [11], mortar specimens [12] and, only in [13]–[15], for actual concrete specimens. Most of the published studies either dealt with model systems as in [12] and/or were not time-lapse [16]–[21]. Only in two studies the ASR cracking was monitored using time-lapse XMT [12], [21]. None of these and of other studies, to the best of our knowledge, has yet investigated (1) the ASR product evolution by XMT (2) in concrete specimens with realistic aggregates. That is because of some intrinsic limitations of XMT.

First, XMT can only identify ASR cracks with thickness larger than the spatial resolution of the acquired tomogram, which is several tens of microns for cm-scale specimens. However, at the early ASR stages, the ASR products precipitate and accumulate within aggregate grain boundaries and micro-cracks, which are intrinsic aggregate features with thickness in part below the spatial resolution of XMT [22] .

Second, later stage products typically appear within cracks of thickness, larger than tens of microns [23] and can in principle be resolved in X-ray tomograms. However, they bear low image contrast to the other material phases, e.g., cement hydrates and aggregate minerals, as shown, for example, in [13], [14]. Such low image contrast stems from similar elemental composition and mass density for products and the other material phases in the aggregates and cement paste. Elemental composition and mass density are the two main properties determining X-ray attenuation, which is at the basis of image contrast in standard X-ray tomography. The lack of X-ray attenuation contrast makes the systematic and reliable identification of the products extremely difficult, limiting, e.g., the successful application of image analysis algorithms for quantitative analysis. As a result of the mentioned issues, it has not yet been possible to take advantage of the non-destructiveness of XMT for unequivocally, reliably and systematically visualizing ASR products.

To overcome this limitation, we propose in this work the addition of Caesium (Cs) in the concrete mix, to serve as a X-ray attenuation contrast enhancer between ASR products and the other material phases, thus *de facto* as a ASR product tracer, for XMT. Leemann and Münch [22] already proposed the addition of Cs to the concrete mix to enhance the visibility of ASR products during scanning electron microscopy. They found that, as an alkali ion, $Cs^+$ becomes naturally incorporated into the ASR products (herein called Cs-doped products) but less so in cement hydrates. Furthermore, Cs does not seem to disturb the ASR mechanisms. What makes Cs a good candidate as a tracer of ASR products also in X-ray tomograms is its atomic number ($Z$ = 55) that is much higher with respect to the other atoms typically present in concrete. That results in larger X-ray photoelectric absorption for the doped products, compared with the other material phases. Furthermore, it has similar hydrated radius as potassium (K), another alkali typically participating in the ASR [22], [24].

With this study, we aimed at assessing the usefulness of Cs-doping for enhancing the non-destructive, time-lapse investigation of ASR and its associated cracking by XMT. We also assessed its influence on the natural ASR evolution by a systematic study on model concrete specimens with and without doping. The specimens' size and mix designs were chosen paying specific attention to achieve a good trade-off between maximizing the tomograms' spatial resolution and allowing the occurrence of meaningful ASR cracking. The latter requires the specimens to contain, along each direction, at least several aggregates with distinct sieve size. This is a necessary condition, since cracking in real world concrete strongly depends upon the mutual interactions between the developing stress fields and several aggregates of distinct size. On the other side, the larger the specimens, the smaller the tomographic spatial resolution, with consequent limitations to the part of the ASR crack networks which can be investigated.

In Section 2, we summarize the XMT measurement settings and the respective 3D image analysis as well, along with other characterization techniques used for comparing specimens with and without doping. The actual details for each measurement type and of the 3D image analysis are provided fully in the Supplementary Data (see appendix A). In Section 3, we report the most significant results. They concern (1) the expansion, induced by ASR, of concrete prisms (2) the morphological and chemical features of the ASR products, (3) some of the features of the spatial-temporal distribution of the ASR products, (4) the spatial-temporal pattern and some of the geometrical features of the ASR crack networks and (5) the spatial-temporal distribution of mechanical deformations associated with such networks. We show how XMT and Cs-doping allow improving the knowledge about the ASR damage chemo-mechanics, by making it possible to track the correlated spatial-temporal (i.e., 4D) distributions of cracks and products. We also show that the proposed image analysis allows a semi-quantitative characterization of such correlated distributions. We conclude in Section 4 with a summary of ASR (damage) mechanisms that were investigated in this study and that could be further studied by combining XMT with Cs-doping.

## 2. EXPERIMENTAL METHODOLOGY

### 2.1 Materials used and time-lapse experimental campaign

The experimental campaign consisted of the following elements:

- concrete specimens with and without Cs were cast, containing reactive aggregates and with size optimized for achieving representativeness and sufficient spatial resolution (by the available X-ray tomograph) to detect a significant part of ASR damage (cracks/products);
- the specimens were subjected to an ASR acceleration protocol in the laboratory;
- at multiple and successive time points during the acceleration, non-destructive measurements on the same exact specimens were performed;
- other destructive measurements, or measurements that might have perturbed the ASR development, were performed on separate specimens at different acceleration times. Here we logically assumed that these separate specimens were representative of the ASR (damage) state in the specimen population.

The specimens were prisms with size 40 mm ($X$-axis), 40 mm ($Y$-axis) and 160 mm ($Z$-axis), respectively. In the following, the $Z$-axis direction is called the longitudinal direction of the specimen. The maximum aggregate sieve size was 11.25 mm. We chose such specimen size and aggregate sieve-size range in order to achieve a trade-off between the representativeness of the volume of concrete undergoing ASR damage and the spatial resolution achievable in the X-ray tomograms. Our laboratory-scale X-ray tomography setup uses a conical X-ray beam. In such case, the maximum achievable spatial resolution is inversely proportional to the maximum lateral specimen size. With our choice of specimen size, we could achieve a spatial resolution of tens of microns, while imaging regions of specimens containing at least several aggregates along each dimension of the volume. The latter feature allowed investigating a concrete volume where the ASR-induced stresses (and respective cracking) could stem from/interact with a sufficient number of aggregates. In concrete structures such interactions, in the presence of many aggregates, strongly drive/affect the cracking patterns.

The specimens were cast with a Portland cement of type CEM I 42.5 N with a $Na_2O$-eq. of 1.26 mass-%. Its chemical composition, measured by X-ray fluorescence spectroscopy (XRF) analysis, was the following (by mass-%): CaO 63.0, $SiO_2$ 20.1, $Al_2O_3$ 4.6, $Fe_2O_3$ 3.3, $SO_3$ 3.3, MgO 1.9, $K_2O$ 0.96, $TiO_2$ 0.37, $P_2O_5$ 0.24, $Na_2O$ 0.16, MnO 0.05, $Cr_2O_3$ 0.01, LOI 2.1. Table 1 below provides the mix composition of the reference specimens, cast without Cs. NaOH was added in an amount leading to a $Na_2O$-eq. of 1.6 mass-%.

The specimens with Cs-doping were cast with an identical mix design (Table 2) except for the addition of $CsNO_3$. In order to keep the $Na_2O$-eq of the concrete the same when compared with the reference concrete, NaOH was added in a smaller amount (Table 2). Such reduction was assumed (and verified by the obtained results shown in Section 3) not to hinder the ASR development, as the added $CsNO_3$ counterbalanced the unavoidable drop in the pore solution alkalinity due to the smaller NaOH content. With the added amount of $CsNO_3$, the alkali molar ratio of the concrete [Cs]/[Na+K] was 0.43.

The choice of Cs as the optimal tracer of ASR products was made in terms of their increased visibility in X-ray tomograms. It was based upon the results of a preliminary experimental campaign which is described in Section S2.2 of the Supplementary Data[1]. In such campaign, distinct specimens doped with distinct alkalis were investigated. As expected, $Cs^+$ ions, incorporated in the ASR products, produced the largest X-ray attenuation contrast in the tomograms, compared with the other alkali ions ($Na^+$, $K^+$ and $Rb^+$). See Figure S7 in the Supplementary Data[2] for a comparison of ASR products labelled with these distinct alkalis, as seen in our X-ray tomograms.

Distinct specimens, with and without Cs-doping, were cast with two types of aggregates, originating from two distinct regions in central (canton Uri, "U" aggregates) and Southwest (Praz, canton Valais, "P" aggregates) Switzerland, respectively. Section S1.1 reports their chemical compositions, obtained by XRF analysis, and their mineralogical composition, obtained by powder X-ray diffraction (PXRD) analysis. The particle density of the U aggregates is 2667 kg · $m^{-3}$ while the particle density of the P aggregates is 2651 kg · $m^{-3}$, both measured according to the EN 1097-6:2000 standard [25]. While both aggregates contained a similar amount of the same minerals, their differences stem from the distribution of those minerals (texture). See Section S1 for the texture characterization results obtained by optical microscopy performed with cross-polarized light (CP-OM), in addition to the chemical and mineralogical characterizations. The U aggregates are sedimentary rocks (impure sandstones) mainly constituted of microcrystalline quartz, including small amounts of amorphous $SiO_2$ (mainly in the form of detrital chert fragments and siliceous inclusions in the cementing intergranular regions, as identified by CP-OM), feldspar and calcite [26]. Foliations, such as white mica flakes, are uniformly distributed in the interstitial regions between fine quartz grain boundaries. The high specific surface area of the small grains along with the presence of detrital chert fragments makes the U aggregates very reactive in the presence of alkalis [27]. The P aggregates on the other hand are metamorphic rocks.

---

[1] Sections of the Supplementary Data document are referred to in what follows shortly as "SX.Y", with X and Y being positive integer numbers, while we drop "Supplementary Data".

[2] In what follows, figures in the Supplementary Data document are referred to as "Figure SZ", with Z a positive integer number, the "S" implying they belong to the Suppl. Data.

Both the specimens with and those without Cs-doping, independently of the aggregate type, were subjected to the same laboratory ASR acceleration protocol. Such protocol was an adaptation of the SIA MB 2042 standard [28], typically used for assessing the degree of ASR reactivity of a certain concrete mix design. The protocol consisted of storing the specimens at 100% relative humidity (RH) and 60 ± 2°C in a climatic chamber for 24 hours immediately after casting. They were then unmoulded and immersed in an alkaline solution (0.3 M KOH, 0.1 M NaOH) contained in boxes. Such a solution simulates the natural pore solution of concrete [29]. The boxes were tightly sealed and stored in an oven with a temperature of 40 °C for up to 250 days. The boundary conditions for the specimens are closer to the natural condition of a field-exposed concrete, compared with those of other acceleration protocols, requiring storage at 60°C or 80°C.

At distinct time points, different specimens were taken out of the sealed boxes to perform various characterisations and measurements on them. The measurements included mass and dimensional change (along the $Z$-axis, also called longitudinal axis of the specimen, see Figure S16), quasi-static Young's modulus along the longitudinal direction (EN 12390-13:2013 standard [30]), flexural strength and compressive strength along the longitudinal direction (both according to the EN 196-1:2016 standard [31]), scanning electron microscopy with back-scatter electron contrast (SEM-BSE), energy-dispersive X-ray spectroscopy (EDX) and XMT. The strength measurements were carried out, at each stage of the campaign, on a distinct batch of three specimens. SEM-EDX was carried out on distinct batches of two specimens. The non-destructive measurements (length and mass changes, quasi-static Young's modulus and XMT) were carried out on the same exact specimens (three in total) all along the campaign. Length and mass measurements were additionally performed also on a second batch of three other specimens. In the case of destructive measurements (SEM-EDX, compressive and flexural strengths), the specimens were randomly selected from additional batches distinct from those used for the non-destructive measurements.

All the measurement types were performed on both the Cs-doped specimens, named in the following as "P-Cs" and "U-Cs" for the two distinct aggregate types, respectively, and the specimens without any doping, used as reference counterparts and named as "P-Ref" and "U-Ref", respectively. The specimens subjected to XMT were measured only up to 250 days. Within such period, the tomographic time series consisted of $N_T$ = 3, 12 and 8 time points for the P-Ref/U-Ref, P-Cs and U-Cs specimens, respectively.

SEM and EDX analyses were used for imaging cracks and ASR products with higher spatial resolution than what achievable by XMT, in order to characterize qualitative features of the ASR damage and the products' morphology and to characterize their chemical composition, respectively. Section S1.2 provides the details about the latter measurements.

We performed standard XMT measurements, whereby, as mentioned in Section 1, X-ray attenuation is the main source of the voxel value contrast. We refer the reader to Section S2.1 for all the details about the settings and properties of these measurements. We report here just the effective spatial resolution of our tomograms, which was approximately 70 $\mu m$ (twice the voxel size of the tomograms).

Table 1. Mix composition of the reference specimens in units of kg·m$^{-3}$ (mass per m$^3$ of cast material).

| Specimen label | Cement CEM I 42.5N | Aggregates | | | Deionized water | NaOH |
|---|---|---|---|---|---|---|
| | | 0 – 4 mm | 4 – 8 mm | 8 – 11.25 mm | | |
| P-Ref/U-Ref | 450 | 659 | 412 | 576 | 225 | 4.9 |

Table 2. Mix composition of the Cs-doped specimens in units of kg·m$^{-3}$ (mass per m$^3$ of cast material).

| Specimen label | Cement CEM I 42.5N | Aggregates | | | Deionized water | NaOH | CsNO$_3$ |
|---|---|---|---|---|---|---|---|
| | | 0 – 4 mm | 4 – 8 mm | 8 – 11.25 mm | | | |
| P-Cs/U-Cs | 450 | 659 | 412 | 576 | 225 | 1.8 | 13.5 |

## 2.2 Analysis of the time-lapse X-ray tomograms

We implemented a 3D image analysis workflow and applied it to one specimen of each type (out of three specimens imaged by XMT) to perform a quantitative analysis which could take advantage of the 4D (space + time) information available. We list and describe in the following the steps of such workflow from a conceptual point of view. The actual image processing details are fully reported in Section S3.

**(I) Image enhancement**

Each tomogram was first of all enhanced, in terms of noise reduction. We used a type of edge-preserving noise filter (see Section S3.1).

**(II) Correction for specimen misalignments**

The tomograms of any specimen, acquired at distinct time points, were not perfectly spatially aligned, because of limitations in the accuracy with which the same specimen can be positioned on the specimen/rotation stage. Any comparative analysis of tomograms in a time series requires first of all reducing such misalignment. For such purpose we applied to the results of step (I) a 3D rigid body (i.e., translation + rotation) registration algorithm (see Section S3.2).

**(III) Crack segmentation**

Crack regions, being either empty or filled by ASR products with very low Cs concentration, were clearly identifiable visually. See Figure 5 for examples. However, image noise and partial volume effects typically make their algorithmic identification (segmentation) strongly dependent on several steps of post-processing to correct for artefacts. Such steps frequently require multiple choices by the image analyst.

In order to reduce the need for such corrections, we implemented an image processing workflow relying on the fact that ASR-generated cracks, as 3D objects, do not exist at the reference time $t_0$ (= 1 day, at the ASR acceleration start) but appear at successive times $t_i$, $\forall\, i = 1, \ldots, N_T$, where $N_T$ as mentioned before, indicates the total number of points in the time series after the reference one. Such approach exploits two, successive image registration steps: global affine followed by non-affine registration. We report in Section S3.3 the details as well as advantages/disadvantages of such workflow and its validation.

This approach has been proposed and implemented in the biomedical image analysis field for detecting, tracking and quantifying changes in, e.g., organ structures [32]. It has been also proposed for locating regions where drying shrinkage-related cracking in concrete occurred [33].

The result of the workflow was an ASR empty crack binary tomogram, i.e., a 3D label map of voxels attributed to ASR-induced crack regions, not containing Cs-labelled ASR products.

Similar crack regions visible already at $t_0$, thus not induced by ASR, were segmented by using a distinct workflow, also described in Section S3.3 and requiring more artefacts reduction.

**(IV) ASR products segmentation**

The accumulation of ASR products took place both in the natural pore space, e.g., in air voids and original cracks[3], and in ASR-induced cracks. Consequently, in Cs-doped specimens part of the ASR cracks that contained products were not segmented as cracks by the workflow described at point (III) above. We describe in Section S3.4 how it was possible to segment pore space, of any type, containing products by still taking advantage of the results of the processing mentioned at point (III). Section S3.4 contains further information about how the ASR products binary tomograms could be partitioned into one for products inside cracks (original or ASR-induced) and another one for products inside other types of pores.

At each time point the binary tomogram for ASR products inside cracks was joined together with the binary tomogram of empty ASR cracks obtained at step (III) to create a final binary tomogram of cracks. This was the binary tomogram analyzed for characterizing the crack network temporal evolution.

**(V) Quantitative characterization of the crack network**

By crack network, we mean here the sets of interconnected clusters of voxels classified by steps (III) and (IV) as belonging to cracks, either original or ASR-induced and either containing ASR products with high Cs concentrations or not.

In addition to computing the volume of such crack network, normalized by the tomogram volume, we computed at each time feature variables of the distinct parts of the crack network to quantitatively characterize the temporal evolution of the network itself. Such quantitative analysis complemented the qualitative one obtained by the 3D rendering of the network's binary tomogram.

We summarize here the definition of some of the feature variables plotted in Section 3. Section S3.5 reports the actual mathematical and computational details of such variables as well as those for other features for which the results are available in the Supplementary Data only.

---

[3] Here and in what follows, the term "original cracks" or "original pores" is used to indicate cracks/pores not generated by ASR, for example cracks existing intrinsically in the aggregates. Such cracks/pores are those detected already in the tomograms at the reference time point $t_0$ = 1 day, before the start of the ASR acceleration protocol.

**Shape tensor analysis**

For each distinct and separated part (also called branch) of the crack network, we computed the parallelepiped box circumscribing the branch (called bounding box in the following) and oriented along its principal axes. The principal axes correspond to the directions of the three *eigen*vectors of the branch's shape tensor $G$. See Section S3.5 for the mathematical definitions of $G$, of its *eigen*value/*eigen*vector analysis and how to exploit such analysis for characterizing size, shape and orientation in space [34]–[38]. We report in the following only the size analysis results. Those for shape and orientation are available in Section S10.1. We used the three lateral size measures of the bounding box as definitions of length ($L$), height (also called width, $H$) and thickness ($T$) of the object, corresponding to the largest, mid and smallest size of the box, respectively ($L \geq H \geq T$) [36].

**Local thickness analysis**

The crack network branch thickness $T$ computed by the shape tensor analysis is a single number which obviously cannot fully take into account the effects on the $T$ value of the complex shape of a crack, especially with increasing degrees of surface folding and tortuosity. In addition to a single, coarse thickness $T$, we therefore also computed a scalar field of local crack thickness, $T_{local}(\vec{x})$. The definition and computational details of such field are presented in Section S3.5.

The visualization of $T_{local}(\vec{x})$ allowed getting qualitative insights into the spatial-temporal distribution of the cracking. The $T_{local}(\vec{x})$ values were assembled together from all the crack network branches to create a whole statistical *ensemble*. We then analyzed the statistics of such ensemble and how it evolved, in order to assess quantitative features of the cracking.

**(VI) Global and local deformations analysis**

One of the outputs of both the global affine and the non-affine registration steps performed for the ASR crack segmentation consisted of a displacement vector field, $\vec{u}(\vec{x}, t_i)$, $\forall\, i = 1, \dots, N_T$. In the global affine registration step, such vector field is linearly dependent in $\vec{x}$ (see Eq. (ES1) in Section S3.3 and Eq. (ES9) in Section S3.6). On the contrary, the respective field associated with the non-affine registration can exhibit strong spatial variations.

We used the global affine displacement vector field to compute the bulk, relative size change,

$$\frac{\Delta L_{AFF,k}}{L_{AFF,k}}(t_i) \equiv \frac{L_{AFF,k}(t_i) - L_{AFF,k}(t_0)}{L_{AFF,k}(t_0)}, \forall\, i = 1, \dots, N_T \tag{1},$$

of the tomographed volume along the axes $k = X, Y$ or $Z$. Section S3.6 provides the details about such computations. On the one side, the comparison of the experimentally measured $\frac{\Delta L_Z}{L_Z}(t_i)$ with the one computed according Eq. (1) allows assessing whether the strain of the tomographed portion of the specimen was the same as the average strain of the whole specimen. In turn, that allows assessing whether the tomographed portion was a representative volume in terms of capturing the overall ASR deformation. On the other side, we could quantify by $\frac{\Delta L_{AFF,k}}{L_{AFF,k}}(t_i)$ the relative change in size also along the $X$ and $Y$ axes. Thus we could compare bulk deformations in directions orthogonal to the specimen's longitudinal direction.

Finally, we used the non-affine registration results to map out local deformations and to investigate their eventual correlations with the ASR cracks and reaction products. For this purpose, we report in Section 3 computation results for the determinant of the Jacobian matrix of the non-affine mapping vector field, indicated with $J_{\vec{T}_{N-AFF}}(\vec{x}, t_i)$. Such scalar field provides a 3D map of local relative volume (local volume at $t_i$ divided by local volume at $t_0$). Values larger than 1 indicate local volumetric expansion while values smaller than 1 local volumetric shrinkage [39].

## 3. RESULTS

We report in this Section only the results for the specimens cast with the Praz (P) aggregates, given overall similarities of results obtained for the specimens cast with the Uri (U) aggregates. The most significant difference consisted of a faster ASR reactivity for the U specimens. The corresponding results for the latter specimen type are reported in the Supplementary Data document (see appendix A).

### 3.1 ASR-induced macroscopic dimensional changes and mechanical properties evolution

The relative length change evolution, both of the whole specimens (experimentally measured) and for the tomographed volumes (measured from the global affine registration results) is shown in Figure 1 (a). Corresponding results for the specimens cast with U aggregates can be found in Section S5.1. All the specimens expanded along the $Z$-axis approximately in a linear fashion but with distinct and slightly different rates. Detailed quantitative analysis of such rates can be found in Section S5.1.

At early age (7 – 14 days), the Cs-doped specimens expanded longitudinally with rates similar to those of their corresponding reference specimens. However, at about 20 days, the doped specimens started to exhibit higher expansion rates. The maximum expansion of the doped specimens was at least about 30% larger than what attained by the corresponding reference specimens. This expansion boosting effect was also observed for the specimens of the preliminary study, by which the doping alkali was chosen and it was already reported in [22]. The other alkalis had qualitatively a similar effect as Cs did. The details about the results from the preliminary study we conducted can be found in Section S4.

As already observed in [22], the main influence of Cs-doping consisted in an increase in ASR kinetics. This result is supported not only by the observed faster expansion but also by a faster mass increase, due to the uptake of the alkaline solution in which the specimens are submerged. A faster/higher uptake correlates well with higher expansion.Cs-doping was accompanied by a faster/higher mass increase, with a maximum value in time about 25 - 40% higher for the doped specimens than for the reference ones. The detailed results about the mass evolution are available in Section S5.2.

The relative length changes of the tomographed volumes (computed as per Section S3.6) differed significantly from the respective changes measured experimentally for the full specimen volume only up to 21 days. See Section S5.1 for the statistical analysis of the relevance of such differences. After 21 days, the minimum and the maximum values of the relative difference between the estimated relative length changes were 0.00042% and 0.0325%, respectively, for all specimen types. This result suggests that, after 21 days, the deformations occurring in the tomographed volume were representative of those at the length scale of the full specimen, despite the fact that only about 30% of the whole specimen volume was tomographed.

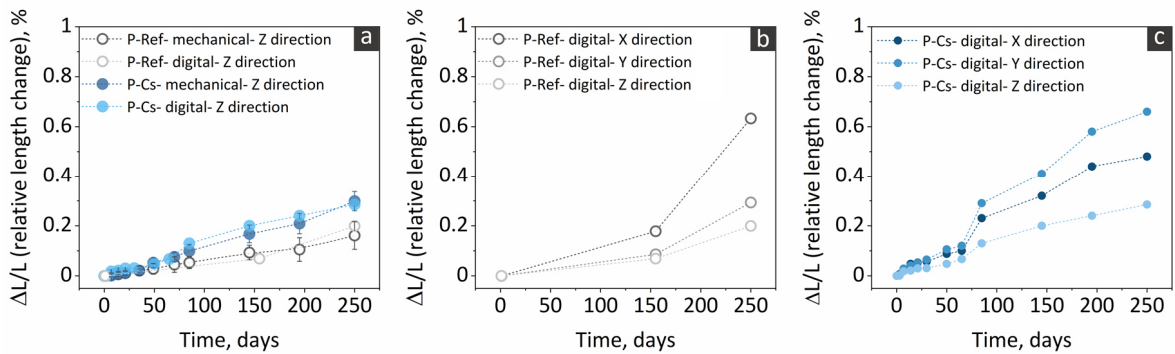

Figure 1. Evolution of the relative dimensional change of the specimens cast with the P aggregates, with (blue shades and filled markers) and without (grey shades and hollow markers) Cs-doping. The label "digital" means the estimate was obtained for the tomographed volume of one specimen of each type, by the procedure described in Section S3.6 ("Global analysis"). The label "mechanical" means the estimate was obtained for the whole specimen volume, by experimental measurements with the gauge described in S3.6 ("Global analysis") as well. (a)

*Results for the relative change in size along the $Z$-axis (i.e., length). The experimentally measured values were obtained, for each mix type, for six specimens (including the one being the subject of the tomography analysis). The marker shows the average of the six values. The error bar shows their standard deviation. (b) and (c) Relative dimensional changes of the tomographed volumes, along the $X$, $Y$ and $Z$ axes for the P-Ref and P-Cs specimens, respectively.*

Figure 1 (b) and (c) compare the relative length changes with the respective relative changes in size of the tomographed volume along the two other axes, $X$ and $Y$, all obtained by the procedure described in Section S3.6. For both the reference and the Cs-doped specimens, the maximum expansion was larger in the lateral directions ($X$ and $Y$) than in the longitudinal direction ($Z$). Similar results for the U specimens are available in Section S5.1.

This difference is likely due to the aspect ratio of the specimens (1:4). Because of the latter, the near-surface zone (with higher moisture and alkali content, thus proner to expansion) contributes to a higher proportion of the size in the $X$ and $Y$ directions than in the $Z$ direction (see also [40], [41]).

The time series of the quasi-static Young's modulus and of the flexural and compressive strengths are shown in Figure 2 (a) to (c), respectively. The temporal evolution of each of these mechanical properties was characterized by an increase up to about 30 to 70 days since the ASR acceleration started, which is attributable to the ongoing cement hydration. Although ASR cracking was likely already occurring during this time period (see the expansion in Figure 1), the positive effect of cement hydration on the mechanical properties was prevalent in this initial period. While both the flexural and the compressive strength peaked at about 70 days and slightly decreased (flexural strength) or remained approximately constant (compressive strength) afterwards, the Young's modulus peaked earlier (about 30 days), after which it monotonically and strongly decreased.

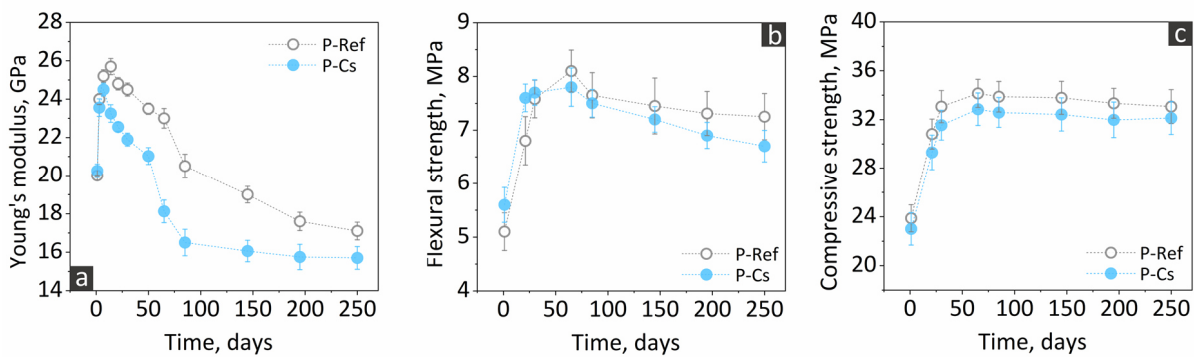

*Figure 2. The time series of the (a) quasi-static Young's modulus, (b) the flexural and (c) the compressive strength of specimens with and without Cs addition, cast with P aggregates.*

The effect of Cs-doping on the mechanical properties was quite clear. The Young's modulus developed at early ages rather similarly, up to the peak. After that, the Cs-doped specimens exhibited a much faster drop than their corresponding reference specimens and achieved a plateau, while the reference specimens' modulus was still decreasing at the end of the measuring campaign. A similar but less remarked effect of Cs-doping is visible in the evolution of the flexural and compressive strengths (Figure 2 (b) and (c), respectively). In summary, the higher levels of expansion due to Cs-doping were mirrored by lower values of macroscopic mechanical properties. Similar results for the U specimens are reported in Section S6.

## 3.2 Chemical and morphological features of the ASR products

Figure 3 (a) and (b) shows SEM-BSE micrographs of ASR products inside cracks from a P-Ref specimen and a P-Cs one, respectively. These micrographs were acquired at 120 days of ASR acceleration time and, for both specimens, from interior regions of the aggregates, where the ASR products are usually rather crystalline [42]. The ASR products developed in the P-Cs case presented a significantly higher image contrast than the ones formed in the P-Ref case, in agreement with what already reported in [22]. As already shown in [22] and [42], in the absence of Cs-doping, the ASR products exhibited similar SEM-BSE pixel value as the surrounding material phases (Figure 3 (a)). Therefore, they could be distinguished only based on their typical morphological features and chemical composition (when performing additional EDX analysis). For crack regions in the aggregate interior, such features consist of plate-like texture [42].

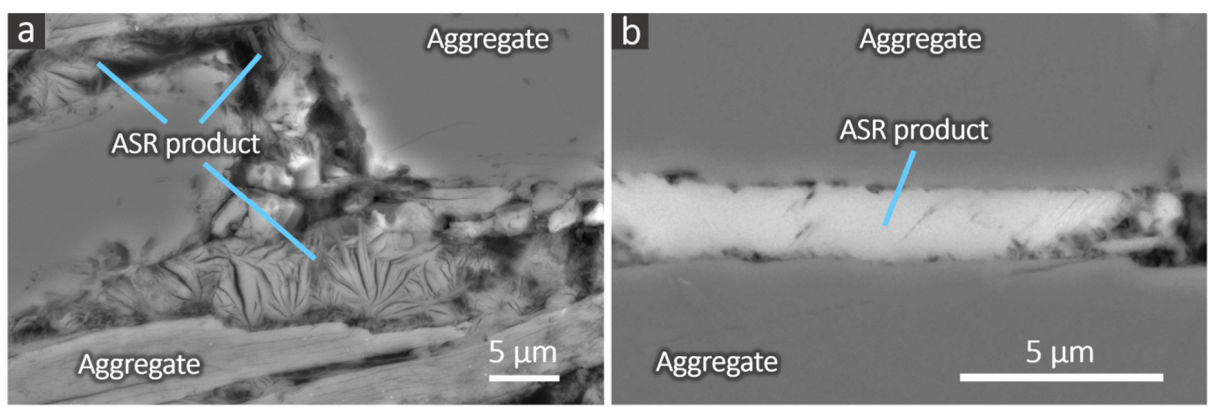

*Figure 3. (a) and (b): examples of SEM (back-scatter electron, or BSE) micrographs of ASR products precipitated inside the cracks of aggregates from a P-Ref specimen and a P-Cs one, respectively.*

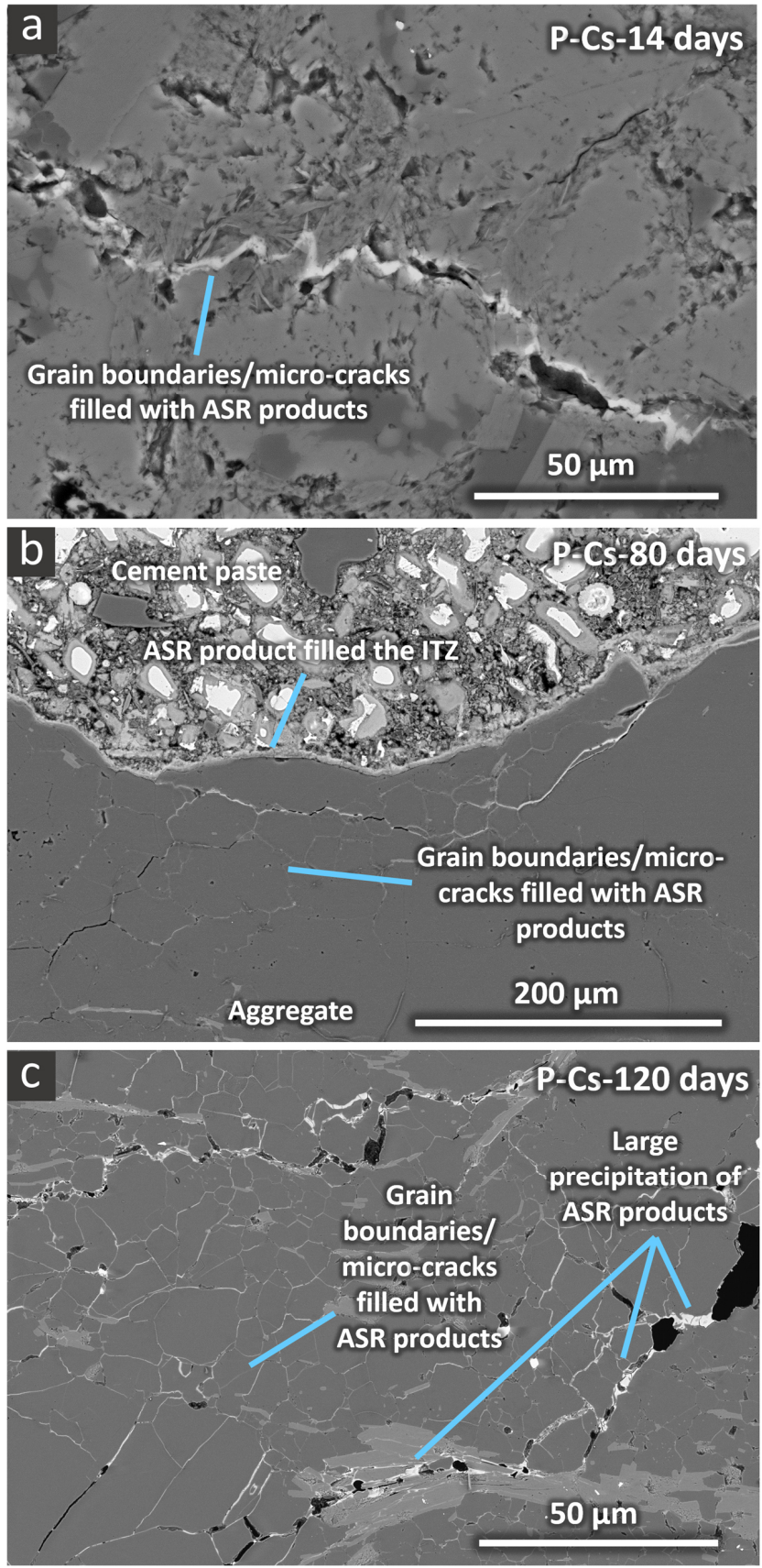

Figure 4. SEM-BSE micrographs providing examples of the distribution of ASR products and cracks in the specimens with Cs, cast with P aggregates. The micrographs were acquired at 14, 80 and 120 days.

In the presence of Cs-doping, the products identification could rely not only on such textural features but also on the significantly enhanced BSE-contrast, confirming that Cs was incorporated in the products. The latter result was confirmed by EDX analysis (see Section S1.2 for the details). Such analysis also showed no evidence of significant compositional differences in the products, with or without embedded Cs, except for its presence.

Cs-doping also allowed confirming the observation, in SEM-BSE micrographs of distinct specimens at distinct time points, of a specific spatial-temporal distribution pattern of ASR products, already reported in [22]. Figure 4 below exemplifies such pattern for P-Cs specimens. Results for the other specimen types are available in Section 1.2.

At 14 days, Cs-containing products could be seen at inter-granular quartz regions close to the aggregate edges and in cracks at the sub-micron and micron scales, in cracked or un-cracked aggregates or in both. Such products are the initial ASR products [22] and could not be detected by XMT, as performed in this study, due to insufficient spatial resolution. At 80 days, more products could be seen in internal regions than close to the edges. This is in agreement with the observations reported in [22] of a sort of ASR reaction/product front moving from the aggregate boundaries towards the inner regions. At 120 days, such front advanced inside the aggregate, while the cracks were on average wider. Although most of the cracks within the aggregates themselves were found empty, some large chunks of products could also be found. They completely filled part of the cracks in the aggregates (highlighted in the micrograph in Figure 4 (c)).

### 3.3 ASR cracking in the reference specimens: time-lapse tomography, qualitative analysis

Due to similar X-ray attenuation, ASR products, aggregates and hydrated cement phases attained very similar voxel values. Thus, the only visible change due to ASR consisted of the cracks. No ASR products within the cracks could be as easily resolved as in the SEM-BSE micrographs of Figure 4. A detailed qualitative analysis can be found in Section S7. In all the tomograms of the reference specimens, we observed that ASR cracks originated always inside the aggregates, then propagated into the cement paste [10], [43]. The cracks also increased, in a step-wise manner, in length and width.

Furthermore, the ASR cracks propagated in some cases starting from defects or cracks that were already evident at 1 day. In addition to being locations of stress localization, such original cracks, grain and sub-grain boundaries inside aggregates have been identified in the literature [22], [44] as loci of enhanced ASR product precipitation and accumulation. Such regions are also more prone to coalesce with propagating cracks. The spatial resolution in our tomograms (about 70 $\mu m$) was not sufficient to resolve actual grain boundaries, which are typically tens of nm in width [22]. Thus, in our tomograms we resolved mainly original cracks.

At any time point, the cracks were distributed rather homogenously in the inspected volumes. This could be observed in the binary tomograms of the segmented cracks. The segmented crack networks appeared to be similar, independently of the aggregate type. The crack networks consisted of rather densely interconnected long and short branches. Further details about qualitative features of the crack networks as discerned from 3D rendering of the binary tomograms are available in Sections S7.

### 3.4 ASR cracking in the Cs-doped specimens: time-lapse tomography, qualitative analysis

Cs-doping clearly allowed tracing the accumulation of ASR products, thus providing important information about the spatiotemporal ASR evolution. This result is exemplified in Figure 5 and Figure 6 for the P-Cs specimen.

Figure 5 shows a 2D cross-section (slice) parallel to the $X-Y$ plane from a 3D region of interest (ROI) of the P-Cs specimen at four different time points. ASR products are evident as very bright regions within cracks and near the aggregate boundaries, especially after 145 days. The aggregate highlighted in Figure 5 (a) contained cracks with size close to the tomographic spatial resolution already at 1 day (original cracks). At 85 days (Figure 5 (b)), the main ASR effect was the filling of pores close to the aggregate boundaries, particularly in the Interfacial Transition Zone (ITZ). After 145 days, the ASR products were extruded towards/into the cement paste via newly created cracks or extremely widened grain boundaries (Figure 5 (c)). In addition to filling in the cracks present in the cement paste on the left-hand side of the central aggregate, the ASR products also filled, at 250 days, the more delaminated ITZ of another adjacent aggregate at its right-hand/bottom side (Figure 5 (d)). At 250 days, it can be also observed that cracks propagated further (top-left corner of Figure 5 (d)), they widened up and formed a continuous crack path by connecting with the other cracks in the cement paste. The movie MS1 in Section S8 shows the complete time series of one slice from a larger ROI containing the one used for Figure 5.

We repeatedly observed in several specimens the extrusion of ASR products from an aggregate (where they were first detected) into the surrounding cement paste, either along a propagating crack or into other pores, e.g., in the ITZ. Cracks that were initially filled with ASR products emptied when the cracks propagated further, while the ASR products were extruded further away in the process.

In some cracks, ASR products were present only on their internal surfaces, while the center was empty. The cracks appeared to have enlarged, with the ASR products splitting and remaining attached to the edges. The opening tip of those cracks was instead filled with a considerable amount of ASR product. See Figure 5 (d) and movie MS1 in Section S8.

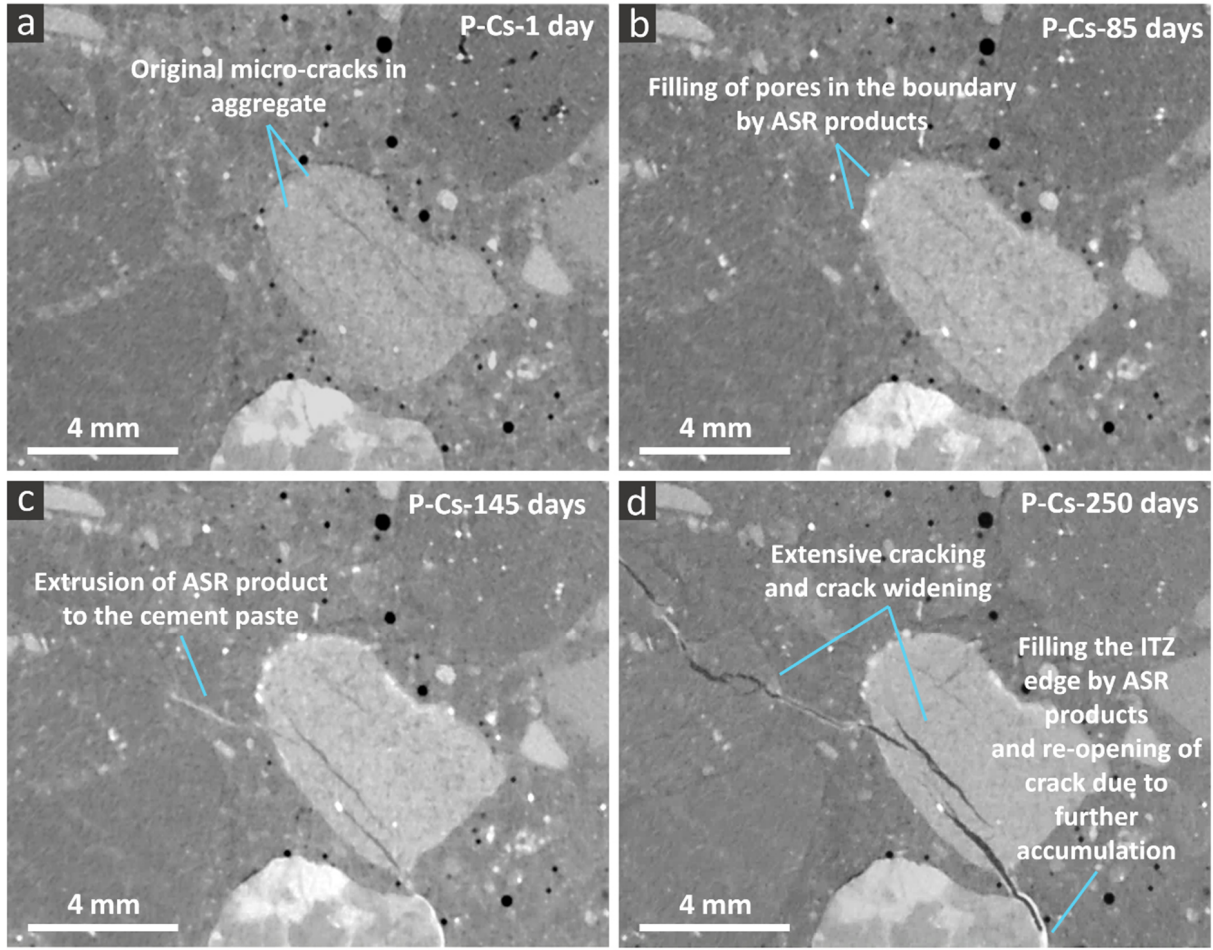

*Figure 5. A slice from a ROI of the P-Cs specimen at four different time points. ASR products can be observed as very bright regions within cracks and near the aggregate boundaries. The four insets here are from the tomogram at (a) 1 day, (b) 85 days, (c) 145 days and (d) 250 days.*

In some cases, the ITZ, original pores and air voids at a distance from a cracking aggregate became gradually filled up with ASR products. Figure 6 shows an example of this observation for a different ROI compared with the one of Figure 5. See the movie MS2 (Section S8) for the full time series of which Figure 6 is only a subset. Figure 6 shows an air void located about 3 mm from an aggregate that got split by one ASR crack. The crack propagated further into the cement paste and through the air void. The crack within the aggregate emptied of ASR prod-ucts and the products were transported along the crack through the cement paste and accumulated into the air void. The accumulation progressed steadily, first with the formation of a meniscus at 65 days. Then, the void became filled at 85 days. Later, at about 145 days, by opening the crack in the cement paste (lower left of air void) part of the ASR gel is depleted and the X-ray attenuation of the remaining products within the voids appeared to increase (indicated by the increase in brightness of the voxels within the void), possibly due to $Ca^{2+}$ ions uptake [45], [46]. At 195 and 250 days, the cracks in the cement paste and aggregate extensively widened. The ASR products extrusion into the cement paste (1) was spatially widespread, (2) it happened in a rather isotropic fashion and not only along the cracks and (3) it led to a widespread filling of pores (not only air voids) in the cement paste around the cracks.

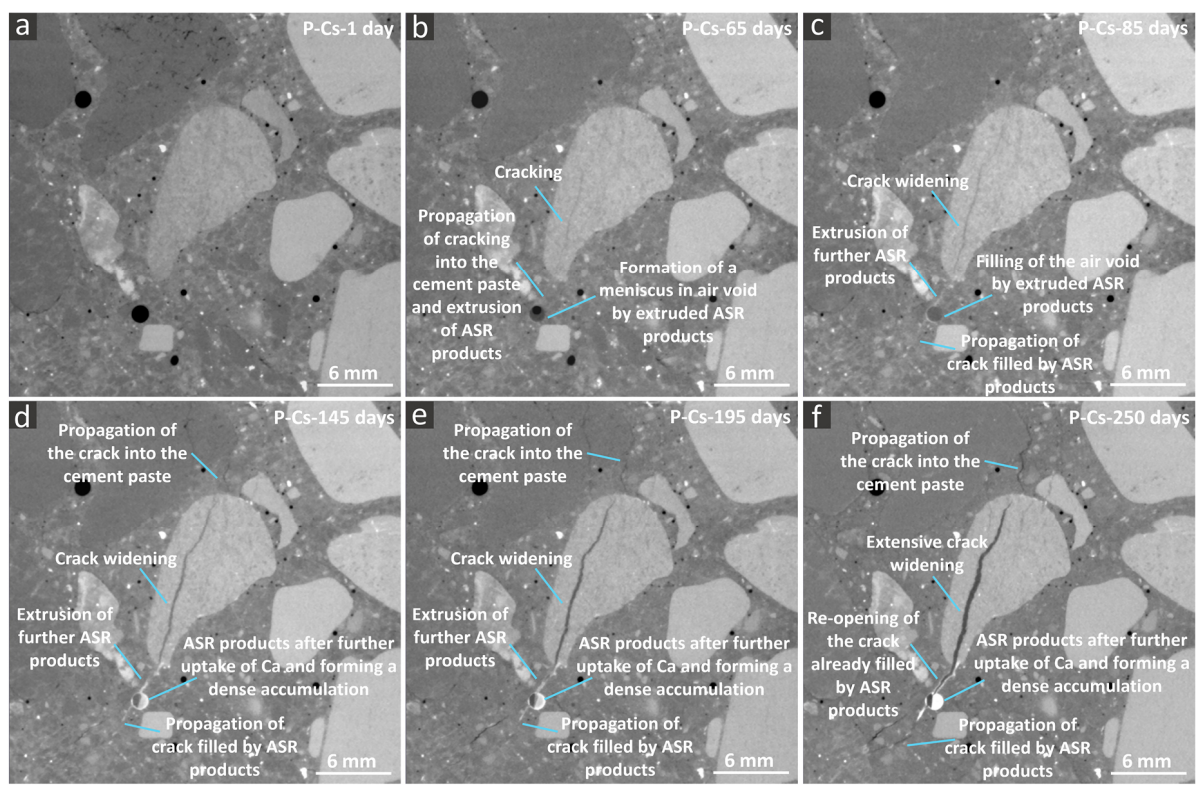

*Figure 6. A slice from another ROI of the P-Cs specimen at four different time points. ASR products could be observed as filling up an air void a few mm away from the aggregate highlighted by the arrow, as a crack running through such aggregate evolved and propagating into the cement paste till the air void itself. The products were transported from such aggregates till the air void via the propagating cracks. The six insets shown here are from the tomogram at (a) 1 day, (b) 65 days, (c) 85 days, (d) 145 days, (e) 195 days and (f) 250 days.*

Figure 7 exemplifies such three features by the rendering, at multiple time points, of a small ROI for two types of binary tomograms: the one of empty cracks and that of ASR products (both inside and outside cracks). Details about the 3D rendering approach are available in Section S8. The chosen ROI contained a single large aggregate. Figure 7 shows that the ASR cracks started their development from original cracks/pores in the aggregate (inset (a)) and gradually propagated with time. The cracks inside the aggregate remained mainly empty, while those in the cement paste became partially or completely filled with products. This can be appreciated mainly at later times (insets (e) and (f)): products regions with the typical features of cracks, e.g., a high degree of shape anisotropy, evolved in distinct parts of the ROI along any possible direction. The small and isolated regions occupied by the products corresponded to other types of pores, e.g., porous patches or ITZ, mainly in the cement paste. Three air voids which became completely filled are observable already at early times (inset (b)). No large crack completely crossed their volumes, as, on the contrary, it happened in the example of Figure 6. Rather, they were in the nearby of crack network branches not propagating towards them. The latter example showcases the second feature mentioned above for the products extrusion, i.e., an isotropic distribution of products through the pore space of the cement paste. Additional qualitative evidence of such and of other features of the products extrusion, as well as of the crack propagation, is provided in Section S8.

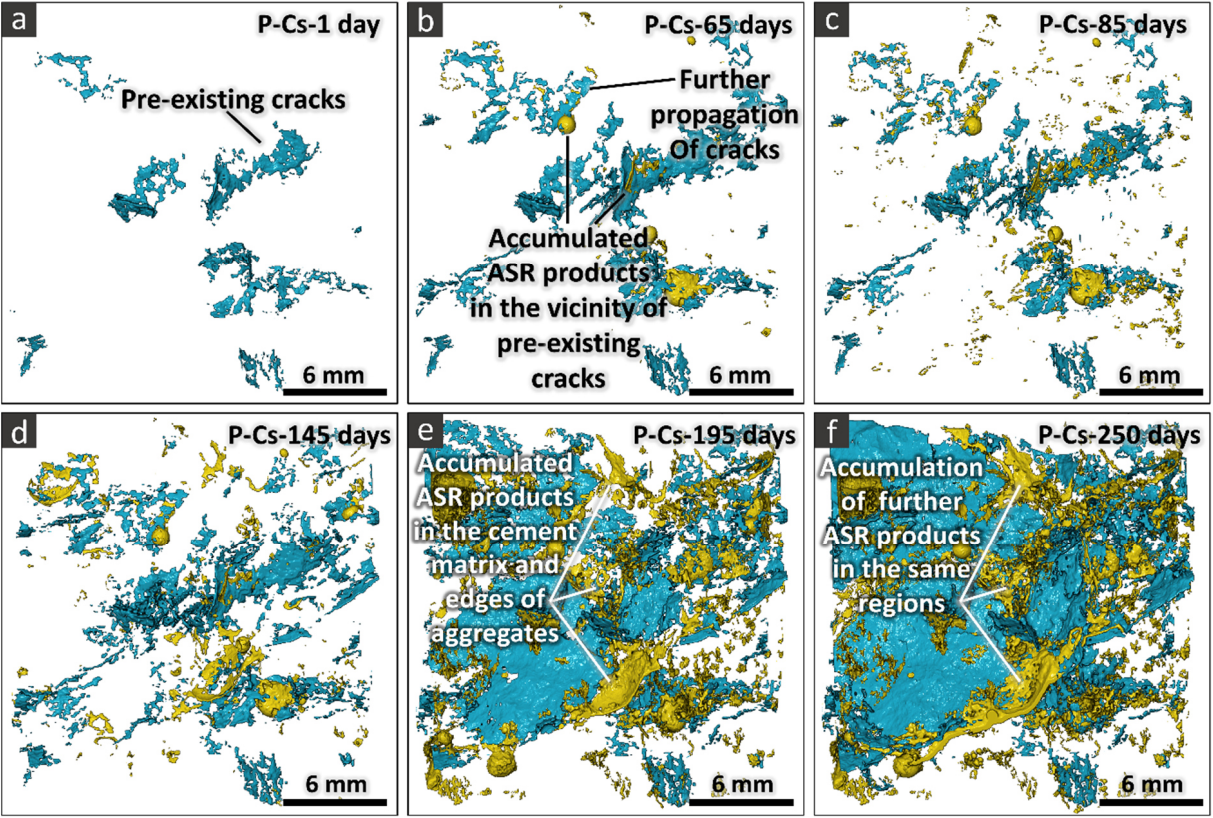

Figure 7. Binary tomograms of the segmented empty cracks (in cyan) and of the ASR products (both inside and outside cracks, in yellow). Only a small ROI of the P-Cs specimen is shown. Such ROI contained cracks that were nucleated inside a single aggregate and propagated later into the surrounding cement paste. The insets from (a) to (f) correspond to different time

*points during the ASR acceleration, including 1 day, 65, 85, 145, 195 and 250 days, respectively.*

### 3.5 ASR crack networks: quantitative analysis

The first quantitative analysis of ASR cracking concerned the evolution of the normalized volume of ASR-generated cracks, defined as

$$V_{ASR\ cracks}(t_i) = \frac{V_{tot,cracks}(t_i) - V_{tot,cracks}(t_0)}{V_{tomogram}} \cdot 100\ [\%], \forall\ i = 1,2,\ldots,N_T \tag{2},$$

where $V_{tot,cracks}(t_0)$ is the total volume of segmented cracks at the reference time point $t_0$, comprising only empty cracks, $V_{tot,cracks}(t_i)$ is the total volume of cracks (empty or filled with ASR products) at a successive time $t_i, \forall\ i = 1,2,\ldots,N_T$, and $V_{tomogram}$ is the tomogram volume. The difference between the first two volumes provides the volume of cracks generated only by the ASR. Expressing such volume with the normalization as *per* Eq. (2) helps when comparing data from tomograms of different size.

Figure 8 shows the $V_{ASR\ cracks}$ time series for both the P-Ref and the P-Cs specimens. It also shows the volume of ASR products (also normalized by $V_{tomogram}$) independently of the type of pore space they occupied (cracks, air-voids, ITZ and porous patches). The comparison of Figure 1 with Figure 8 shows a feature which we observed independently of the specimen type: a positive correlation between the normalized volume of ASR-generated cracks and the longitudinal expansion. The P-Cs specimens achieved, at 250 days, almost double longitudinal expansion compared with the P-Ref (see Figure 1). The tomographed P-Cs specimen exhibited almost three times larger volume of ASR-generated cracks (see Figure 8). Additional details about such positive correlation can be found in Section S9.

Figure 8 also shows that the total volume of ASR products (inside or outside cracks) evolved synchronously with the total volume of ASR cracks and was at most about 16% smaller than it (at 250 days). With the possibility of distinguishing between products inside and outside cracks (see Section 2.2.IV), we could compute the fraction of filled ASR crack volume, which achieved a maximum value of 30% for the tomographed P-Cs specimen. Section S9 reports the full time series of such fraction of filled ASR crack volume. These results mirror what observable in the 3D renderings of the same type as the one shown in Figure 7: a significant accumulation of ASR products outside of the ASR cracks. This is in agreement with the qualitative observation mentioned before of extensive detection of products in pores other than cracks, the majority of which located in the cement paste.

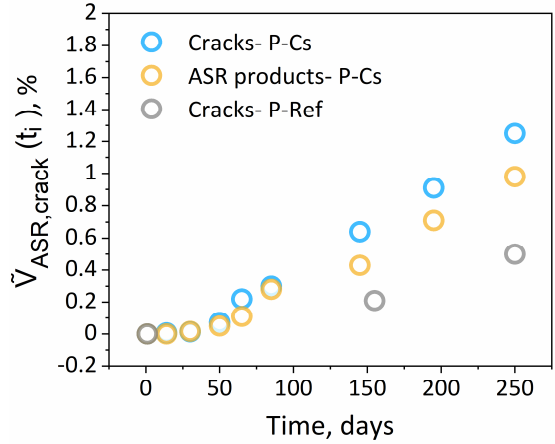

*Figure 8. Quantitative assessment of the ASR cracking in the P specimens by the total volume of ASR-generated cracks (with or without ASR products) normalized by the tomogram volume (named as $V_{ASR\ cracks}$). The total volume of ASR products (also normalized by the tomogram volume) is also shown by yellow markers.*

Concerning the ASR crack size analysis, of the three size definitions obtained from the shape tensor ($G$) analysis, i.e., the bounding box's length $L$, its height or width $H$ and its thickness $T$ (Section 2.2.V), the latter did not provide meaningful estimates. That is because of the characteristic geometrical features of the crack network branches as 3D objects. As mentioned in Sections 3.3 and 3.4, they had high aspect ratios and an indicative shape of thin, curved and fragmented plates. This feature was independent of (1) the aggregate type, (2) the presence or absence of Cs-doping and (3) of time. The detailed results the latter statement is based upon are provided in Section S10.1.

Such curved shape impaired the meaningfulness of the $T$ values. On the contrary, $L$ maintained a high degree of representativeness for the maximum size of a crack. The detailed confirmation of such results, based on the shape tensor analysis, as well as the qualitative (i.e., visual) and quantitative assessments of the reliability and meaningfulness of the $L$, $H$ and $T$ values are provided in Section S10.1 as well. As explained in Section 2.2.V, instead of the $T$ values, we relied on the statistics of the computed local thickness field $T_{local}(\vec{x})$ for the quantitative analysis of crack thickness.

Figure 9 shows the distributions of $L$ and of $T_{local}(\vec{x})$ values for the P-Ref (insets (a) and (c)) and P-Cs (subfigures (b) and (d)) specimens. Only 3 time points from the whole time series are shown for both the P-Ref and the P-Cs specimens, chosen such that they are at similar times as for P-Ref specimen. The distributions at all points of the P-Cs time series are available in Section S10.1 and Section 10.2 for $L$ and $T_{local}(\vec{x})$, respectively. The distributions are shown in the form of Zipf's plots [47]–[49]. Such plot type shows the statistical *ensemble* estimate of the complementary cumulative distribution function (cCDF), $G_X(x)$, of a random variable $X$, in $\log_{10}$-$\log_{10}$ scales. For $X = L$, each separate crack contributed to one sampled value of $L$. For $X = T_{local}$, each voxel inside the crack network provided a sampled value. The choice of reporting Zipf's plots, for both variables, stemmed from the fact that both variables, for each specimen and at any time point, exhibited right-skewed and long-tailed probability distribution functions (PDFs). These two features lead, in a Zipf's plot, to a line (perfect power law PDF) or to one which deviates from a line for a slight amount of downward (exponential or stretched exponential/Weibull or gamma PDF) curvature. A Zipf's plot is one of the first tools in the recognition and analysis of such PDFs [48], [50], [51]. On the one side it allows magnifying any eventual useful feature of the long tail of the PDF. On the other side, it avoids the nuisances with the optimal choice of bin size and choice of the bandwidth(s) for traditional histograms and for kernel density estimators of PDFs, respectively [48], [52].

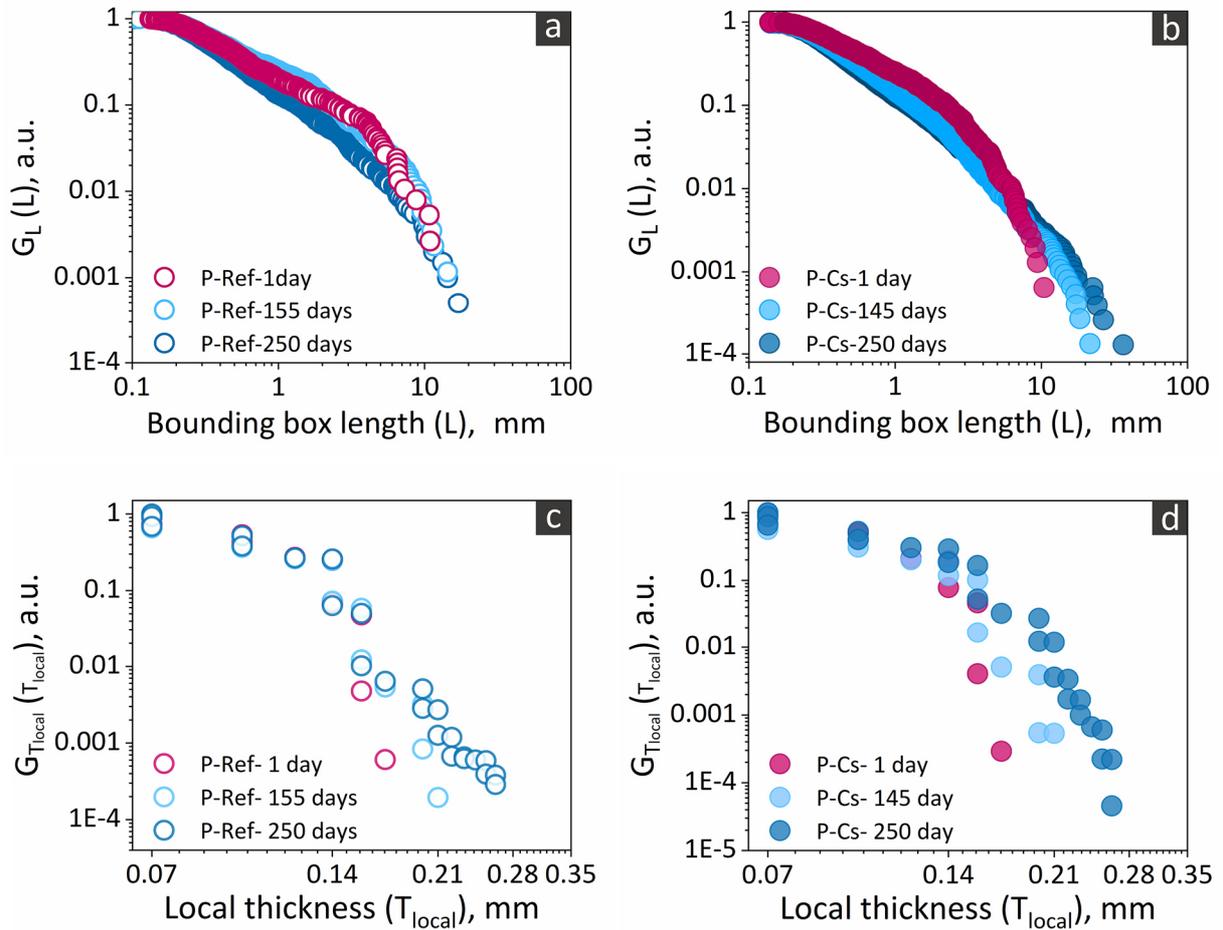

*Figure 9. Empirical (i.e., statistical ensemble) complementary cumulative distribution function of the bounding box length, $L$, $G_L(L)$, in $\log_{10}$-$\log_{10}$ scales (Zip'f plot). Each separate crack contributed to one sampled value for $L$.*

The first important observation is that right-skewed and long-tailed PDFs, as we observed for $L$ and $T_{local}$ are to be expected to govern the distribution of crack size values (especially length and thickness), for cracking in brittle materials. Additional results, showing long-tailed PDFs for other crack size metrics ($H$ and the crack network branch volume, $V_{crack}$), are reported in Section S10.1. Such types of PDFs are typically found for feature variables of processes characterized by criticality and by (certain degrees of) scale invariance [53]. The latter is a property of geometrical features of crack networks in brittle materials. Such property has been subject of intense debates and investigations [54]. The fact that $L$ and $T_{local}$ exhibited right-skewed and long-tailed distributions at any time during the cracking process and, more importantly, independently of Cs-doping, indicates the absence of spurious effects on the ASR cracking by the doping. Similar results for the U specimens are available in Section S10.1. The consequences of ASR kinetics acceleration by Cs-doping significantly concerned the tails of the distributions of $L$ and $T_{local}$. For example, at 250 days, the maximum $L$ values reached about 40 mm for the P-Cs specimen (Figure 9 (b)) while only about 20 mm for the P-ref one (Figure 9 (a)). Similar results obtained for the U specimens are provided in Section S10.1.

Overall, independently from the absence or the presence of Cs-doping, the ASR cracking was accompanied by more significant changes to the overall distribution of $T_{local}$ than to that of $L$. In Figure 9, clearly detectable temporal changes are indeed observable only at the very end of the right side tail of the $L$ distribution. On the contrary, cracking brought a more evident shift of the overall $T_{local}$ distribution towards larger values. $T_{local}(\vec{x})$, as a scalar field, showed, at any given time point, a rather homogenous distribution. Detailed results of the spatial analysis of $T_{local}(\vec{x})$ are reported in Section S10.2.

In terms of crack orientation, we observed no influence of Cs-doping and of the aggregate type. All the specimens showed a slight preferential tendency to be the most elongated along the $Z$ axis. We refer to Section S10.1 for the detailed and quantitative analysis of crack orientation based upon the shape tensor analysis implemented as explained in Section S3.5. We notice that a preferential elongation in the $Z$ direction is in good agreement with the results about the macroscopic expansion along the three axes, as estimated based upon the affine registration procedure. We recall from Figure 1, insets (b) and (c), that the expansion along the $X$ and $Y$ axes were larger than or at least equal to the one along the $Z$ axis. Section S5.1 reports corresponding and similar results for the U specimens.

### 3.6 Local deformations analysis

In order to characterize the cracking in the distinct specimens we visualized, at distinct time points, the determinant of the Jacobian matrix, $J_{\vec{T}_{N-AFF}}(\vec{x}, t_i)$, associated with the transformation vector field of the final non-affine registration step, $\vec{T}_{N-AFF}(\vec{x}, t_i)$. See Section S3.6 for its exact definition. We show the results for the P specimens here, focusing on features which were commonly observed also for the U specimens. The results for the U specimens, as well as additional results based upon a second local deformation metric, are shown in Section S11. The $J_{\vec{T}_{N-AFF}}(\vec{x}, t_i)$ scalar field is shown, at specific time points, in Figure 10 and Figure 11 for the P-Ref and P-Cs specimens, respectively. In each of these figures, only one slice from the specimen's tomographed volume is shown (insets in the first row or first two rows). The slice at the same position but from the 3D distribution of $J_{\vec{T}_{N-AFF}}$ values is rendered, in the last row, semi-transparently according to a color scale and overlapped on top of the respective tomogram's slice.

Figure 10 (d) and (e) showcase some features of the ASR-induced deformations that were observed as systematically accompanying cracking, independently of the aggregate type. First of all, regions where newly formed ASR cracks appeared were systematically characterized by local expansion. These are the yellow to red regions in Figure 10 (d) and (e). There, $J_{\vec{T}_{N-AFF}}$ achieved values larger than one, corresponding to local volumetric increase compared with the reference time point $t_0$. With time, the local expansion became larger and also covered a larger area around the crack. This was consistently observed for all the newly formed cracks. An enlargement of zones of high expansion (marked in red) in the $J_{\vec{T}_{N-AFF}}$ map far away from the cracks was observed in particular for cracks that opened up (see, e.g., the cracks pointed at by lines in Figure 10 (d)-(e), features (I) and (II)). Regions of local volumetric contraction (blue/cyan in Figure 10 (d)-(e), with $J_{\vec{T}_{N-AFF}}$ values smaller than 1) were systematically observed to evolve, both in spatial extent and in intensity, in between regions of local volumetric expansion. The observed alternation of expansion zones to contraction ones, with high degree of localization of extreme values (minima and maxima of $J_{\vec{T}_{N-AFF}}$), is to be expected from cracking in concrete, because of its high degree of spatial heterogeneity/disorder (both in terms of physical properties and geometrical/size ones of its distinct material phases) and its brittleness.

In Figure 10, another notable feature is evident: some original cracks or more porous regions at the aggregate boundaries did not evolve into an opening and lengthening crack, rather they either remained the same or gradually closed or filled up. See as examples the features labelled as (III) in Figure 10. The $J_{\vec{T}_{N-AFF}}$ typically had in the proximity of such regions values smaller than or equal to 1 (Figure 10 (d)-(e)), confirming that local volumetric contraction occurred at these locations. According to our observations, ASR cracking did not always nucleate and advance from any observed original crack or highly porous region. Rather some of these regions even contracted due to the spatially heterogeneous distribution of deformations accompanying the ASR cracking. Our observations cannot however exclude that ASR cracks may nucleate and propagate preferentially from original cracks or highly porous regions that are smaller than 70 $\mu m$, the spatial resolution in our tomograms.

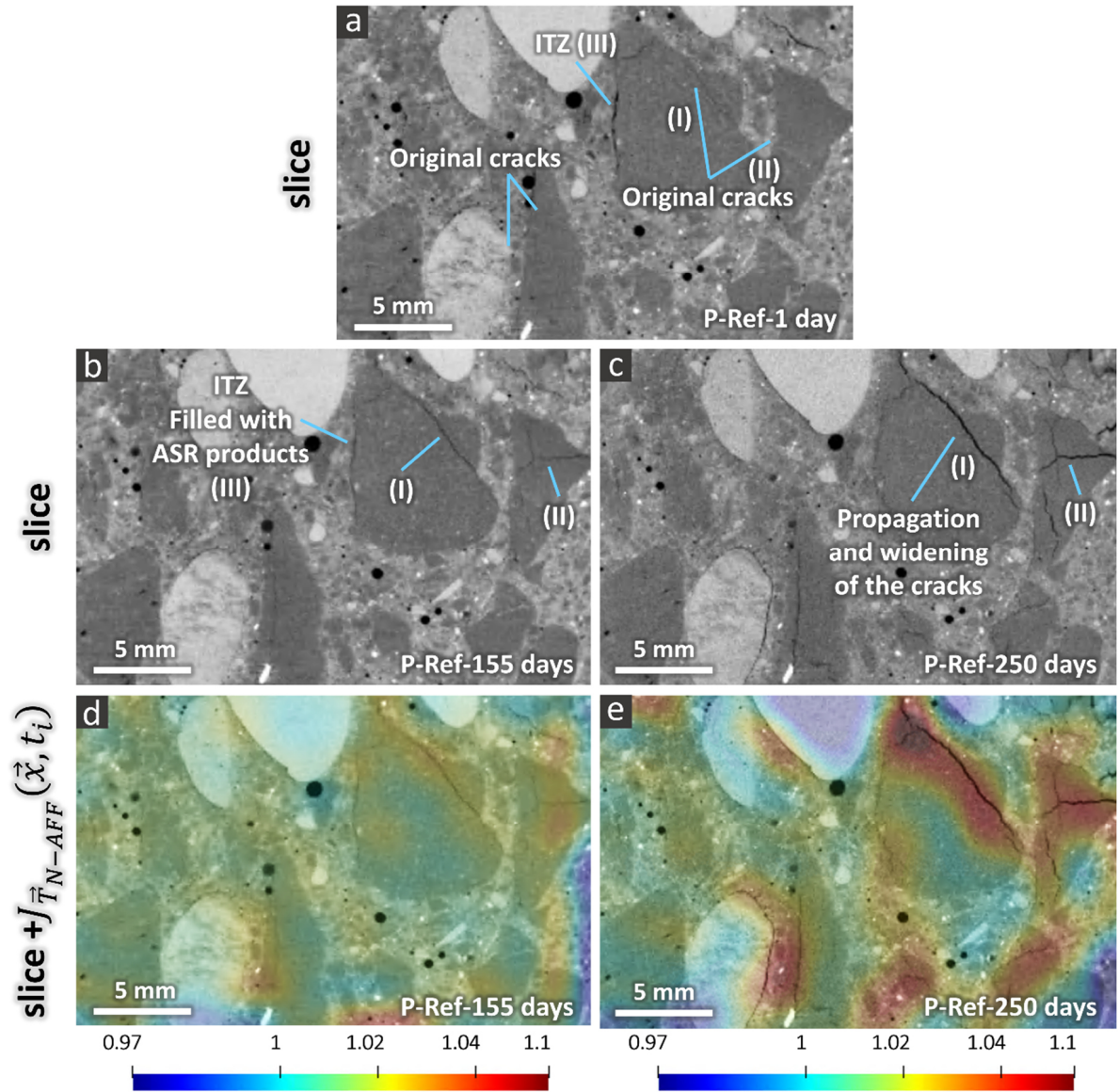

Figure 10. Visualization of the determinant of the Jacobian matrix ($J_{\vec{T}_{N-AFF}}(\vec{x}, t_i)$) of the transformation vector field $\vec{T}_{N-AFF}(\vec{x}, t_i)$ associated with the non-affine registration. This scalar field is used as a spatial map of the factor by which the volume locally expanded or contracted due to the spatially heterogeneous components of the ASR-induced displacement vector field. In this figure, only one 2D cross-section (slice) from the tomographed volume is shown: (a) X-ray tomogram of the P-Ref specimen at the beginning of the ASR acceleration (at 1 day); (b) and (c) slices from the X-ray tomograms at 155 and 250 days, respectively; (d) and (e) the same slices as in (b) and (c), plus, overlapped on top of them semi-transparently and rendered according to the indicated color scale, the 2D cross-section, at the same position, from $J_{\vec{T}_{N-AFF}}(\vec{x}, t_i)$, at the same corresponding time points. The scale bars of insets (d) and (e) have *arbitrary units since $J_{\vec{T}_{N-AFF}}$ is a dimensionless variable. Values greater than 1 represent volumetric expansion, while values smaller than 1 indicate volumetric contraction.*

The development of regions of local volumetric expansion exactly where ASR cracks appeared was a feature systematically observed also for Cs-doped specimens. The evolution of such regions followed temporal patterns which were similar to those observed for the reference specimens. The first yellow/red spots (where $J_{\vec{T}_{N-AFF}} > 1$) appeared in the Cs-doped specimen already after 30 days (Figure 11 (e)). However, such first expansion hotspots appeared when no ASR crack could be actually resolved in the corresponding tomograms. These zones of localized expansion might precede ASR cracking or might indicate that cracks are already present but smaller than the spatial resolution of the tomograms.

With time, the expansion regions increased both in spatial extent, all around the respectively developing ASR cracks, and in intensity (i.e., $J_{\vec{T}_{N-AFF}}$), see Figure 11 (f)-(h). Regions of high local contraction also appeared around those of local expansion. Thus, from a qualitative point of view, the reference and the Cs-doped specimens exhibited similar spatial-temporal $J_{\vec{T}_{N-AFF}}$ maps. The [minima, maxima] values of $J_{\vec{T}_{N-AFF}}$, for the Cs-doped specimen and at 250 days, reached higher values ([0.8, 1.12]) compared to those achieved by the reference specimen ([0.8, 1.09]).

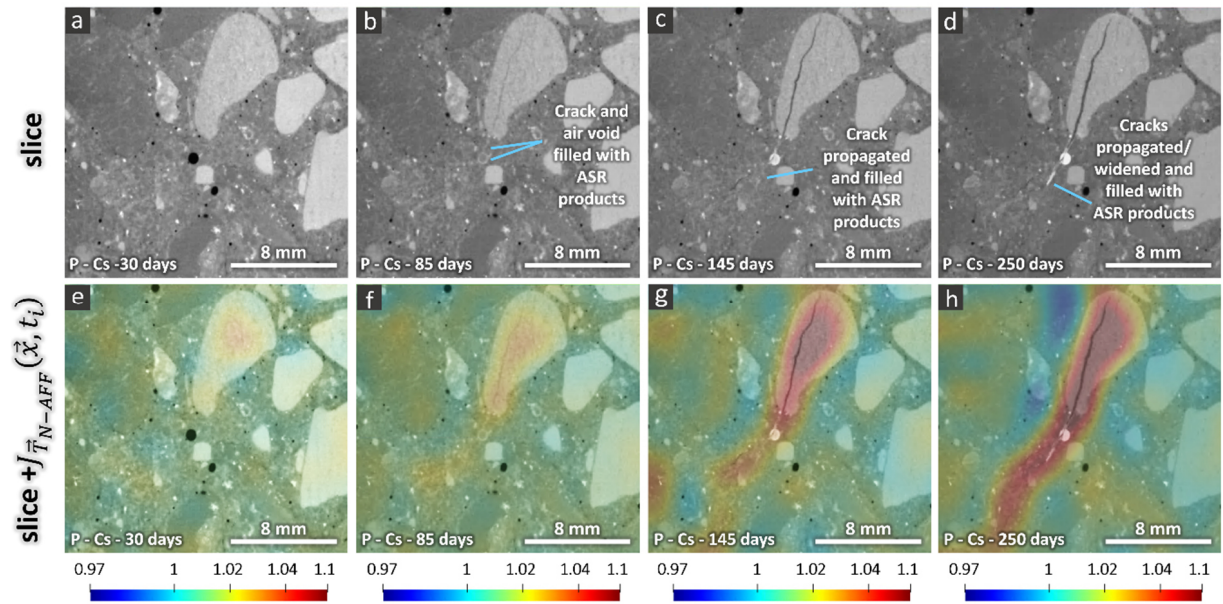

*Figure 11. Similar plots as in Figure 10 but for the P-Cs specimen. The chosen time points for the visualization are 30 days (insets (a) and (e)), 85 days ((b) and (f)), 145 days ((c) and (g)), 250 days ((d) and (h)). The first time point of visualization is here 30 days, instead of 1 day as*

*in Figure 10. At such age, no ASR cracks could be resolved. Thus, the corresponding registered tomogram was essentially identical to the one at 1 day, providing a view of the reference cracking state. However, local deformations appeared already at 30 days, which motivates showing the $J_{\vec{T}_{N-AFF}}(\vec{x})$ scalar field already at such time point.*

## 4. DISCUSSIONS

Time-lapse XMT allowed observing the ASR cracking of the exact same regions in each specimen. This is necessary for achieving reliable and meaningful understanding of the cracking process and for formulating hypotheses about the cracking mechanisms. XMT and the respective 3D image analysis also allowed comparing the ASR cracking on a quantitative basis, i.e., by the use of statistics of specific feature variables. No significant differences due to the Cs-doping were detected for the cracking (Figure 9) and for the spatial-temporal patterns of the associated local deformations (Figure 10 and Figure 11), except for time lags due to the faster ASR kinetics. This study thus provides extensive evidence, complementary to that in [22], that Cs-doping can be used for the visualization of ASR products by, e.g., electron microscopy or X-ray tomography without perturbing the ASR (damage) evolution in laboratory-cast and ASR-accelerated specimens. Further assessment of the Cs-doping usefulness in ASR research could involve specimens cast with distinct aggregate types covering a broader spectrum of ASR reactivity degrees than what covered in our study.

More importantly, the results presented in this paper provided several useful insights into general features of ASR and its associated cracking. We remark that we observed most of these features not only in the specimens with Praz aggregates, whose results are reported in Section 3, but also in the specimens with the Uri aggregates, whose results are reported only in the Supplementary Data (see Appendix A). In the following, we summarize these features and contextualize them within the existing knowledge framework about the chemo-mechanics of ASR. We also provide perspectives on how our work could be extended in the future to advance the understanding of ASR cracking.

It is known that ASR starts with the dissolution of reactive silica in the aggregates [55]. The dissolved silica lead to the formation of an alkali-silica sol [46] . When the concentration in the sol reaches supersaturation ASR products are formed. The silica dissolution occurs at no preferential location in the aggregates. However, at the early ages, formation of ASR products occurs first at the periphery of the aggregate, close to the cement paste, where alkali and $Ca^{2+}$ ions are more available. With time, they progressively precipitate from the boundaries to the interior of the aggregates, forming a so called advancing ASR front [22]. This process occurs thanks to the transport of the mentioned ions along grain boundaries and original cracks, as observed both in our SEM-BSE images for the early ASR products (Figure 4 (b)) and in those reported by [22].

Therefore, the associated early stage cracking is also expected to advance from the boundaries to the center of the aggregates. However, the size of the early-stage cracks are much smaller than the spatial resolution of our tomograms, as confirmed by the SEM-BSE images. Higher spatial resolution (i.e., tens/hundreds nm length scale) X-ray tomography on Cs-doped specimens may be extremely useful for actually observing sequentially, on the same exact specimen, the spatial-temporal pattern of the ASR reaction front and the associated early cracking. Until now, the very early cracking stages have only been observed in 2D, non-time-lapse datasets, i.e., SEM observations on distinct specimens at distinct times [22].

Our observations based on XMT suggest that large amounts of ASR products accumulated preferentially at few specific locations at the aggregate boundaries. For example, we could systematically observe that large (>70 $\mu m$) original cracks or delaminated ITZs were systematically filled up with ASR products (see the examples of delaminated ITZs in Figure 5, top-left corner of the central aggregate). This can be easily understood considering (1) the higher likelihood of high $Ca^{2+}$ concentration in such regions and (2) the fact that such regions could act as reservoirs where transported products can accumulate. With ongoing time, such accumulation can lead to widening of the cracks and their further propagation, even into the cement paste (see Figure 5 (c) and (d) and the associated movie MS1 in Section S8, which shows more time frames than what shown in Figure 5).

In addition, we observed that, along some cracks, the ASR products were transported at distances of up to several mm, as reported in Section 3.4. Both Figure 5 and Figure 6 (and the respective Movies MS1 and MS2 in Section S8 which complement them) show two clear examples of such process, occurring gradually up to a certain point at later times. The ASR products were systematically observed to accumulate in cement paste cracks, ITZs and natural air voids in the cement paste. Such an accumulation, even at several mm from the aggregate where the products formed, can happen due to the extrusion of ASR gel (as a less viscous phase) outside of the aggregates. The occurrence of such extrusion and the corresponding filling of, e.g., air voids, even up to several mm away from ASR-cracking aggregates, have been already inferred before from observations in optical and electron micrographs reported in the literature [56]–[58]. However, to our knowledge, ours are the first ever reported direct, time-lapse and non-destructive *in situ* 3D observations of such transport process.

Figure 6 and the associated movie MS2 in Section S8 provides a clear visual evidence of the described mm-scale transport and accumulation of real-world ASR products. There, the bottom air void was reached, after 55 days, by a crack stemming from the aggregate on its top-right side and propagating through the cement paste. ASR products moved through the evolving crack and gradually accumulated in the air void (compare the voxel value increase inside the air void between 75 and 250 days, which stems partly from an increase in volumetric concentration of products). Additional products reached the air void via the connecting crack and accumulated on top of the older products. Simultaneously, the average voxel value of the previously accumulated products increased in time and was larger than that of the newly accumulating products (compare the frame at 150 days with the one at 200 days in the movie MS2). The latter feature could be explained as an effect of $Ca^{2+}$ ions absorption by the already accumulated products [59]. With such absorption, the mechanical properties (including the rheological ones) of the products are expected to evolve. Evidence of a key role for $Ca^{2+}$ ions in the build-up of stiffness (as quantified by the storage and loss shear moduli) of forming, laboratory synthesized analogs of ASR gels was originally provided by Gaboriaud *et al.* by shear rheometry measurements [60], [61]. More recent experimental work, also on laboratory synthesized analogs [62], has additionally confirmed an hypothesis [45] according to which, also at later ages, the products' stiffness may increase as a consequence of $Ca^{2+}$ ions absorption exchanged with alkali ions (so called alkali recycling mechanism [46]). Based upon such results, reported in the literature, and what observable, for example, in Figure 6 and in the respective movie MS2, it can be concluded that the products transport characteristics and its spatial-temporal extent may be strongly influenced by the local supply of $Ca^{2+}$ ions and the corresponding evolution of the products' mechanical properties. We thus notice that taking into account couplings between $Ca^{2+}$ ions transport/uptake and temporal changes in the products' mechanical properties may end up being a key and necessary ingredient of computational models of ASR cracking working at the microstructure and mesostructure length scales. Although there is in the literature lack of data about mechanical properties, e.g., elastic moduli [23], [63], [64] and viscosity [64], of ASR products extracted from real world concrete, the data measured for their laboratory synthesized analogs and reported in [62] constitute a good starting point for the development of fully chemo-mechanical computational models taking into account reactive transport of $Ca^{2+}$ ions.

We observed that the ASR product transport and accumulation occurred more frequently towards the outer boundary of the aggregates. This could be explained, on the one side, by the tendency of the silica-reach, thick solution and of the alkalis to become inter-dispersed more homogeneously. Due to the presence of a concentration gradient (more silica-reach solution within the aggregate and more alkalis and $Ca^{2+}$ in the cement paste), the silica-reach solution moves out and forms stiffer products by $Ca^{2+}$ uptake. On the contrary, alkalis move inwards (i.e., within the cracks in aggregates) and lead to further silica dissolution. On the other side, products formed and accumulated at the aggregate boundaries may be already stiffer than those formed more inside the aggregate, because of the higher likelihood of $Ca^{2+}$ ions uptake at the boundaries than inside.

Despite the products accumulation in cracks, both inside and outside the aggregates, aggregate dissolution and product accumulation continued (Figure 5 and Figure 6). Further accumulation was frequently accompanied by widening of the cracks and breakdown of the previously formed ASR products. This was evidently observed both by SEM (Figure 4) and by XMT (for example, Figure 5 (d)) results, in the form of broken ASR products sticking to the inner surfaces of widened cracks. Thus, complete crack filling and clogging by products was not frequently observed. Such result is confirmed quantitatively by the computed ASR crack filling ratio, whose maximum value was 30% for the P-Cs specimen. The new products usually further precipitated/accumulated within the same cracks, which became wider.

The spatial-temporal maps of $J_{\vec{T}_{N-AFF}}$ shown in Section 3.6 (Figure 10 and Figure 11) indicate the existence of spatial correlations between localized deformations, mainly expansion, and cracks. In addition, as mentioned in Section 3.6, we observed, for certain aggregates, localization of expansion (Figure 11) before crack appearance, which might indicate the presence of cracks smaller than the spatial resolution of the tomograms. An increase in both the temporal and the spatial resolutions of tomography might help not only with investigating the ASR reaction front and the associated early cracking mentioned above. It might also allow investigating the causal relationships between the measured local deformations and the resolved cracking. The temporal resolution increase is mainly limited by instrument access. On the contrary, the spatial resolution increase is more fundamentally limited by the intrinsic, negative correlation with the specimen size (higher spatial resolution is achievable only for smaller specimens).

Local tomography [65] may help, for cm-scale specimens as used in our work, to improve the spatial resolution by a factor of two to three. However, that will not suffice for resolving early stage products accumulation and respective initial crack propagation. The latter targets could be resolved using the approaches presented in this work together with acquiring datasets with tomographic microscopes based upon synchrotron radiation. The latter measurements would be feasible only for extremely simplified model systems whose realization is extremely challenging and whose representativeness of ASR in actual concrete would need to be carefully assessed and proven.

## 5. CONCLUSIONS

In this study, time-lapse X-ray micro-tomography (XMT), along with doping of specimens with caesium (Cs), was adopted to gain a unique insight into the spatial-temporal evolution of ASR cracking. Cs was used as contrast agent to make ASR products recognizable. Specimens cast without Cs were also used to assess any eventual perturbation by Cs to the ASR product formation and respective cracking.

We could confirm that the incorporation of Cs into the ASR products leads to significant increase of their X-ray attenuation, which becomes higher than for all other phases in the cement paste and in the aggregates. Thus the ASR products can be made easily detectable by image processing algorithms.

We analyzed quantitatively the spatial-temporal patterns of both ASR products and of the respective crack networks as well as the accompanying local deformations. By combining such analysis with macroscopic measurements of dimensional changes and of mechanical properties, we could conclude that the use of Cs as a tracer of ASR products does not appear to perturb the ASR process and the associated cracking, except for accelerating its kinetics.

In this study, at a length scale range down to 70 $\mu m$, we could already take advantage of such methodology to observe and characterize the following features of the coupled ASR products formation - ASR cracking.

a) ASR cracks were first generated in the aggregates and later propagated into the cement paste. Along with crack propagation, we observed, for the first time in a time-lapse manner, the extrusion of ASR products into the cement paste.

b) The extrusion suggests that, at early stages, the ASR products may be a low-viscosity gel which can move away from the originating aggregate and can accumulate later elsewhere as a more stiff phase upon uptake of Ca. Progressive accumulation of ASR products within cracks and air voids in the paste, with following Ca uptake from them, thus their stiffening, was observed in this work. Successive extrusions of low-viscosity, lower Ca-content products being added in time were also observed.

c) The accumulation of ASR products in air voids and cracks in the cement paste or in the ITZ of aggregates surrounding those from which the products originated was observed systematically. Significant portions of the ASR-generated crack volume remained at later ages empty of ASR products.

In the future, the workflow for 3D image analysis and product tracing developed in this work could be directly implemented with XMT characterized by higher spatial resolution for investigating the early stages of ASR products formation and respective cracking. This could be done, e.g., by exploiting local tomography approaches or by using synchrotron radiation-based XMT on smaller model systems. This could help with shedding light on some of the chemo-mechanical coupled phenomena at the basis of the early ASR cracking.

## CRediT authorship contribution statement

**Mahdieh Shakoorioskooie**: Conceptualization, Methodology, Software, Validation, Formal Analysis, Investigation, Data Curation, Writing – Original Draft, Visualization.

**Michele Griffa**: Conceptualization, Methodology, Software, Validation, Formal Analysis, Data Curation, Writing – Original Draft, Visualization, Resources, Supervision, Project Administration, Funding Acquisition.

**Andreas Leemann**: Conceptualization, Methodology, Resources, Writing – Review & Editing, Supervision, Project Administration, Funding Acquisition.

**Robert Zboray**: Conceptualization, Methodology, Resources, Writing – Review & Editing, Supervision.

**Pietro Lura**: Conceptualization, Methodology, Resources, Writing – Review & Editing, Supervision, Funding Acquisition.

## Declaration of competing interest

The authors declare no conflicts of interest.

## Acknowledgments

This work was funded by the Swiss National Science Foundation (SNF), Sinergia project Nr. 171018 (http://p3.snf.ch/project-171018, "Alkali-silica reaction in concrete"). The 3D image analysis was performed by the use of the Empa Platform for Image Analysis, maintained at Empa's Center for X-ray Analytics. We thank the support of Boris Ingold with the preparation of the specimens for the SEM/EDX analysis. We thank Dr. Frank Winnefeld for conducting the XRD/XRF analysis.

## Appendix A. Supplementary data.

Supplementary data to this article can be found online at:
https://doi.org/10.5281/zenodo.4815512

# References

[1] "DIN EN 196-2 : 2013 | METHOD OF TESTING CEMENT - PART 2: CHEMICAL ANALYSIS OF CEMENT | SAI Global." [Online]. Available: https://infostore.saiglobal.com/en-us/Standards/DIN-EN-196-2-2013-386164_SAIG_DIN_DIN_877353/. [Accessed: 11-Nov-2020].

[2] B. Fournier and M.-A. Bérubé, "Alkali-aggregate reaction in concrete: a review of basic concepts and engineering implications," *Can. J. Civ. Eng.*, vol. 27, no. 2, pp. 167–191, 2000, doi: 10.1139/cjce-27-2-167.

[3] A. Leemann, E. Menéndez, and L. Sanchez, "Assessment of Damage and Expansion," in *RILEM State-of-the-Art Reports*, vol. 31, Springer Science+Business Media B.V., 2021, pp. 15–40.

[4] K. Barbotin, S., Boehm-Courjault, E., Leemann, A., & Scrivener, "Characterization of initial ASR products by SEM, FIB and STEM-EDX," in *Peterson & O. Chernoloz (Eds.), 17th euroseminar on microscopy applied to building materials- University of Toronto*, 2019, pp. 141–147.

[5] A. Leemann, Z. Shi, and J. Lindgård, "Characterization of amorphous and crystalline ASR products formed in concrete aggregates," *Cem. Concr. Res.*, vol. 137, p. 106190, Nov. 2020, doi: 10.1016/j.cemconres.2020.106190.

[6] B. P. Gautam and D. K. Panesar, "The effect of elevated conditioning temperature on the ASR expansion, cracking and properties of reactive Spratt aggregate concrete," *Constr. Build. Mater.*, vol. 140, pp. 310–320, Jun. 2017, doi: 10.1016/j.conbuildmat.2017.02.104.

[7] H. Kagimoto, Y. Yasuda, and M. Kawamura, "Mechanisms of ASR surface cracking in a massive concrete cylinder," *Adv. Concr. Constr.*, vol. 3, no. 1, pp. 39–54, Mar. 2015, doi: 10.12989/acc.2015.3.1.039.

[8] Z. P. Bažant, G. Zi, and C. Meyer, "Fracture Mechanics of ASR in Concretes with Waste Glass Particles of Different Sizes," *J. Eng. Mech.*, vol. 126, no. 3, pp. 226–232, Mar. 2000, doi: 10.1061/(asce)0733-9399(2000)126:3(226).

[9] K. F. Hanson, T. J. Van Dam, K. R. Peterson, and L. L. Sutter, "Effect of Sample Preparation on Chemical Composition and Morphology of Alkali–Silica Reaction Products," *Transp. Res. Rec. J. Transp. Res. Board*, vol. 1834, no. 1, pp. 1–7, Jan. 2003, doi: 10.3141/1834-01.

[10] E. Boehm-Courjault, S. Barbotin, A. Leemann, and K. Scrivener, "Microstructure, crystallinity and composition of alkali-silica reaction products in concrete determined by transmission electron microscopy," *Cem. Concr. Res.*, vol. 130, p. 105988, Apr.

2020, doi: 10.1016/j.cemconres.2020.105988.

[11] S. Guo, Q. Dai, X. Sun, and X. Xiao, "X-ray CT characterization and fracture simulation of ASR damage of glass particles in alkaline solution and mortar," *Theor. Appl. Fract. Mech.*, vol. 92, pp. 76–88, Dec. 2017, doi: 10.1016/j.tafmec.2017.05.014.

[12] S. Yang, H. Cui, and C. S. Poon, "Assessment of in-situ alkali-silica reaction (ASR) development of glass aggregate concrete prepared with dry-mix and conventional wet-mix methods by X-ray computed micro-tomography," *Cem. Concr. Compos.*, vol. 90, pp. 266–276, Jul. 2018, doi: 10.1016/j.cemconcomp.2018.03.027.

[13] F. Weise, K. Voland, S. Pirskawetz, and D. Meinel, "Innovative measurement techniques for characterising internal damage processes in concrete due to ASR," in *Proceedings of the International Conference on Alkali Aggregate Reaction (ICAAR), University of Texas, Austin, TX, USA*, 2012, pp. 20–25.

[14] F. Weise, K. Voland, S. Pirskawetz, and D. Meinel, "Analyse AKR-induzierter Schädigungsprozesse in Beton: Einsatz innovativer Prüftechniken," *Beton- und Stahlbetonbau*, vol. 107, no. 12, pp. 805–815, Dec. 2012, doi: 10.1002/best.201200049.

[15] P. Leemann, A., Borchers, I., Shakoorioskooie, M., Griffa, M., Müller, C., & Lura, "MICROSTRUCTURAL ANALYSIS OF ASR IN CONCRETE-ACCELERATED TESTING VERSUS NATURAL EXPOSURE," in *Proceedings of the International Conference on Sustainable Materials, Systems and Structures (SMSS2019), RILEM Publications SARL, Rovinj, Croatia*, 2019, pp. 222–229.

[16] S. Guo, Q. Dai, X. Sun, X. Xiao, R. Si, and J. Wang, "Reduced alkali-silica reaction damage in recycled glass mortar samples with supplementary cementitious materials," *J. Clean. Prod.*, vol. 172, pp. 3621–3633, Jan. 2018, doi: 10.1016/j.jclepro.2017.11.119.

[17] N. Marinoni, M. Voltolini, L. Mancini, P. Vignola, A. Pagani, and A. Pavese, "An investigation of mortars affected by alkali-silica reaction by X-ray synchrotron microtomography: A preliminary study," *J. Mater. Sci.*, vol. 44, no. 21, pp. 5815–5823, Nov. 2009, doi: 10.1007/s10853-009-3817-9.

[18] N. Marinoni *et al.*, "A combined synchrotron radiation micro computed tomography and micro X-ray diffraction study on deleterious alkali-silica reaction," *J. Mater. Sci.*, vol. 50, no. 24, pp. 7985–7997, Dec. 2015, doi: 10.1007/s10853-015-9364-7.

[19] M. Voltolini, N. Marinoni, and L. Mancini, "Synchrotron X-ray computed microtomography investigation of a mortar affected by alkali-silica reaction: A quantitative characterization of its microstructural features," *J. Mater. Sci.*, vol. 46, no. 20, pp. 6633–6641, Oct. 2011, doi: 10.1007/s10853-011-5614-5.

[20] T. Kim, M. F. Alnahhal, Q. D. Nguyen, P. Panchmatia, A. Hajimohammadi, and A. Castel, "Initial sequence for alkali-silica reaction: Transport barrier and spatial distribution of reaction products," *Cem. Concr. Compos.*, vol. 104, p. 103378, Nov. 2019, doi: 10.1016/j.cemconcomp.2019.103378.

[21] D. Hernández-Cruz, C. W. Hargis, J. Dominowski, M. J. Radler, and P. J. M. Monteiro, "Fiber reinforced mortar affected by alkali-silica reaction: A study by synchrotron microtomography," *Cem. Concr. Compos.*, vol. 68, pp. 123–130, Apr. 2016, doi: 10.1016/j.cemconcomp.2016.02.003.

[22] A. Leemann and B. Münch, "The addition of caesium to concrete with alkali-silica reaction: Implications on product identification and recognition of the reaction sequence," *Cem. Concr. Res.*, vol. 120, pp. 27–35, Jun. 2019, doi: 10.1016/j.cemconres.2019.03.016.

[23] A. Leemann and P. Lura, "E-modulus of the alkali-silica-reaction product determined by micro-indentation," *Constr. Build. Mater.*, vol. 44, pp. 221–227, Jul. 2013, doi: 10.1016/j.conbuildmat.2013.03.018.

[24] E. R. Nightingale, "Phenomenological theory of ion solvation. Effective radii of hydrated ions," *J. Phys. Chem.*, vol. 63, no. 9, pp. 1381–1387, 1959, doi: 10.1021/j150579a011.

[25] "CEN, Prüfverfahren für mechanische und physikalische Eigenschaften von Gesteinkörnungen – Teil 6: Bestimmung der Rohdichte und der Wasseraufnahme, EN 1097-6:2001 D." [Online]. Available: https://www.beuth.de/de/norm/din-en-1097-6/34244218. [Accessed: 07-May-2021].

[26] C. Bärtschi, "Kieselkalke der Schweiz: Charakterisierung eines Rohstoffs aus geologischer, petrographischer, wirt-schaftlicher und umweltrelevanter Sicht," *PhD Thesis, ETH Zurich*, 2011, doi: 10.3929/ETHZ-A-006471191.

[27] I. Fernandes, M. A. T. M. Broekmans, M. dos A. Ribeiro, and I. Sims, "Sedimentary Rocks," in *Petrographic Atlas: Characterisation of Aggregates Regarding Potential Reactivity to Alkalis*, Dordrecht: Springer Netherlands, 2016, pp. 43–101.

[28] "SIA MB 2042, Vorbeugung von Schäden durch die Alkali-Aggregat-Reaktion (AAR) bei Betonbauten, Schweizerischer In-genieur- und Architektenverein (2012)." [Online]. Available: https://www.sia.ch/de/dienstleistungen/artikelbeitraege/detail/article/neues-merkblatt-sia-2042/.

[29] B. Lothenbach and F. Winnefeld, "Thermodynamic modelling of the hydration of Portland cement," *Cem. Concr. Res.*, vol. 36, no. 2, pp. 209–226, Feb. 2006, doi: 10.1016/j.cemconres.2005.03.001.

[30] "CEN, Prüfung von Festbeton – Teil 13: Bestimmung des Elastizitätsmoduls unter Druckbelastung (Sekantmodul), EN 12390-13:2013 D." [Online]. Available: https://www.beuth.de/de/norm/din-en-12390-13/187189769. [Accessed: 11-Jan-2021].

[31] "DIN EN 196-1: 2016 DE - Test method for cement - Part 1: Determination of strength; German version EN 196-1: 2016 (Foreign Standard)." [Online]. Available: https://webstore.ansi.org/standards/din/dinen1962016de. [Accessed: 11-Jan-2021].

[32] R. Sakamoto *et al.*, "Temporal subtraction of serial CT images with large deformation diffeomorphic metric mapping in the identification of bone metastases," *Radiology*, vol. 285, no. 2, pp. 629–639, Nov. 2017, doi: 10.1148/radiol.2017161942.

[33] F. Hild *et al.*, "ON THE USE OF 3D IMAGES AND 3D DISPLACEMENT MEASUREMENTS FOR THE ANALYSIS OF DAMAGE MECHANISMS IN CONCRETE-LIKE MATERIALS ON THE USE OF 3D IMAGES AND 3D DISPLACEMENT MEASUREMENTS FOR THE ANALYSIS OF DAMAGE MECHANISMS IN CONCRETE-LIKE MATERIALS. VIII," in *International Conference on Fracture Me-chanics of Concrete and Concrete Structures*, 2013.

[34] M. R. Teague, "IMAGE ANALYSIS VIA THE GENERAL THEORY OF MOMENTS.," *J. Opt. Soc. Am.*, vol. 70, no. 8, pp. 920–930, Aug. 1980, doi: 10.1364/JOSA.70.000920.

[35] D. N. Theodorou and U. W. Suter, "Shape of Unperturbed Linear Polymers: Polypropylene," *Macromolecules*, vol. 18, no. 6, pp. 1206–1214, 1985, doi: 10.1021/ma00148a028.

[36] S. T. Erdoğan, E. J. Garboczi, and D. W. Fowler, "Shape and size of microfine aggregates: X-ray microcomputed tomography vs. laser diffraction," *Powder Technol.*, vol. 177, no. 2, pp. 53–63, Aug. 2007, doi: 10.1016/j.powtec.2007.02.016.

[37] M. Zemp, O. Y. Gnedin, N. Y. Gnedin, and A. V. Kravtsov, "On determining the shape of matter distributions," *Astrophys. Journal, Suppl. Ser.*, vol. 197, no. 2, p. 30, Dec. 2011, doi: 10.1088/0067-0049/197/2/30.

[38] H. Goldstein, C. P. Poole, and J. L. Safko, "The Rigid Body Equations of Motion: Classical Mechanics, 3rd ed.," *Pearson*, 2013. .

[39] G. B. Arfken, H. J. Weber, and F. E. Harris, "Vector Analysis in Curved Coordinates and Tensors: Mathemat-ical Methods for Physicists, 6th ed.," *Elsevier Academic Press*, 2005. .

[40] C. Merz and A. Leemann, "Assessment of the residual expansion potential of concrete from structures damaged by AAR," *Cem. Concr. Res.*, vol. 52, pp. 182–189, Oct. 2013, doi: 10.1016/j.cemconres.2013.07.001.

[41] J. Lindgård, E. J. Sellevold, M. D. A. Thomas, B. Pedersen, H. Justnes, and T. F.

Rønning, "Alkali-silica reaction (ASR) - Performance testing: Influence of specimen pre-treatment, exposure conditions and prism size on concrete porosity, moisture state and transport properties," *Cem. Concr. Res.*, vol. 53, pp. 145–167, Nov. 2013, doi: 10.1016/j.cemconres.2013.05.020.

[42] A. Leemann, "Raman microscopy of alkali-silica reaction (ASR) products formed in concrete," *Cem. Concr. Res.*, vol. 102, pp. 41–47, Dec. 2017, doi: 10.1016/j.cemconres.2017.08.014.

[43] P. Rivard, B. Fournier, and G. Ballivy, "The damage rating index method for ASR affected concrete - A critical review of petrographic features of deterioration and evaluation criteria," *Cem. Concr. Aggregates*, vol. 24, no. 2, pp. 81–91, Dec. 2002, doi: 10.1520/cca10531j.

[44] C. Rößler, B. Möser, C. Giebson, and H. M. Ludwig, "Application of Electron Backscatter Diffraction to evaluate the ASR risk of concrete aggregates," *Cem. Concr. Res.*, vol. 95, pp. 47–55, May 2017, doi: 10.1016/j.cemconres.2017.02.015.

[45] S. Urhan, "Alkali silica and pozzolanic reactions in concrete. Part 1: Interpretation of published results and an hypothesis concerning the mechanism," *Cem. Concr. Res.*, vol. 17, no. 1, pp. 141–152, Jan. 1987, doi: 10.1016/0008-8846(87)90068-8.

[46] F. Rajabipour, E. Giannini, C. Dunant, J. H. Ideker, and M. D. A. Thomas, "Alkali-silica reaction: Current understanding of the reaction mechanisms and the knowledge gaps," *Cem. Concr. Res.*, vol. 76, pp. 130–146, Jul. 2015, doi: 10.1016/j.cemconres.2015.05.024.

[47] P. Cirillo, "Are your data really Pareto distributed?," *Phys. A Stat. Mech. its Appl.*, vol. 392, no. 23, pp. 5947–5962, Dec. 2013, doi: 10.1016/j.physa.2013.07.061.

[48] M. Newman, "Power laws, Pareto distributions and Zipf's law," *Contemp. Phys.*, vol. 46, no. 5, pp. 323–351, Sep. 2005, doi: 10.1080/00107510500052444.

[49] G. K. Zipf, *Human Behaviour and the Principle of Least Effort*. Reading, MA: Addison-Wesly, 1949.

[50] A. Clauset, C. R. Shalizi, and M. E. J. Newman, "Power-law distributions in empirical data," *SIAM Review*, vol. 51, no. 4. Society for Industrial and Applied Mathematics, pp. 661–703, 06-Nov-2009, doi: 10.1137/070710111.

[51] D. Sornette, "Probability Distributions in Complex Systems," *Encycl. Complex. Syst. Sci.*, pp. 7009–7024, 2009, doi: 10.1007/978-0-387-30440-3_418.

[52] Z. I. Botev, J. F. Grotowski, and D. P. Kroese, "Kernel density estimation via diffusion," *Ann. Stat.*, vol. 38, no. 5, pp. 2916–2957, Nov. 2010, doi: 10.1214/10-AOS799.

[53] D. Sornette, *Critical Phenomena in Natural Sciences*. Berlin, Heidelberg: Springer

Berlin Heidelberg, 2000.

[54]	E. Bonnet *et al.*, "Scaling of fracture systems in geological media," *Rev. Geophys.*, vol. 39, no. 3, pp. 347–383, Aug. 2001, doi: 10.1029/1999RG000074.

[55]	G. Davies and R. E. Oberholster, "Alkali-silica reaction products and their development," *Cem. Concr. Res.*, vol. 18, no. 4, pp. 621–635, Jul. 1988, doi: 10.1016/0008-8846(88)90055-5.

[56]	T. Knudsen and N. Thaulow, "Quantitative microanalyses of alkali-silica gel in concrete," *Cem. Concr. Res.*, vol. 5, no. 5, pp. 443–454, Sep. 1975, doi: 10.1016/0008-8846(75)90019-8.

[57]	A. D. Jensen, S. Chatterji, P. Christensen, and N. Thaulow, "Studies of alkali-silica reaction - part II effect of air-entrainment on expansion," *Cem. Concr. Res.*, vol. 14, no. 3, pp. 311–314, May 1984, doi: 10.1016/0008-8846(84)90046-2.

[58]	V. Jensen, "Alkali-aggregate reaction in Southern Norway," *Dr. thesis, Tech. Univ. Trondheim, NTH, Norw.*, p. pp 265 +10 Appendices, 1993.

[59]	M. D. A. Thomas, "The role of calcium hydroxide in alkali recycling in concrete, in: Materials Science of Concrete Special Volume on Calcium Hydroxide in Concrete," *Spec. Vol. Calcium Hydroxide Concr. J. Skaln. J. Gebauer, I. Odler, Eds., Am. Ceram. Soc. Westerv. (OH), USA*, pp. 269–280, 2001.

[60]	F. Gaboriaud, A. Nonat, D. Chaumont, and A. Craievich, "Aggregation Processes and Formation of Silico-calco-alkaline Gels under High Ionic Strength," *J. Colloid Interface Sci.*, vol. 253, no. 1, pp. 140–149, Sep. 2002, doi: 10.1006/JCIS.2002.8522.

[61]	F. Gaboriaud, A. Nonat, D. Chaumont, and A. Craievich, "Structural model of gelation processes of a sodium silicate sol destabilized by calcium ions: combination of SAXS and rheological measurements," *J. Non. Cryst. Solids*, vol. 351, no. 4, pp. 351–354, Feb. 2005, doi: 10.1016/J.JNONCRYSOL.2004.11.019.

[62]	A. Gholizadeh-Vayghan, F. Rajabipour, M. Khaghani, and M. Hillman, "Characterization of viscoelastic behavior of synthetic alkali–silica reaction gels," *Cem. Concr. Compos.*, vol. 104, p. 103359, Nov. 2019, doi: 10.1016/j.cemconcomp.2019.103359.

[63]	G. Geng *et al.*, "Mechanical behavior and phase change of alkali-silica reaction products under hydrostatic compression," *Acta Crystallogr. B. Struct. Sci. Cryst. Eng. Mater.*, vol. 76, no. 4, pp. 674–682, Aug. 2020, doi: 10.1107/S205252062000846X.

[64]	C. Zhang, L. Sorelli, B. Fournier, J. Duchesne, J. Bastien, and Z. Chen, "Stress-relaxation of crystalline alkali-silica reaction products: Characterization by micro- and nanoindentation and simplified modeling," *Constr. Build. Mater.*, vol. 148, pp. 455–464,

Sep. 2017, doi: 10.1016/J.CONBUILDMAT.2017.05.069.

[65] A. Kyrieleis, V. Titarenko, M. Ibison, T. Connolley, and P. J. Withers, "Region-of-interest tomography using filtered backprojection: Assessing the practical limits," *J. Microsc.*, vol. 241, no. 1, pp. 69–82, Jan. 2011, doi: 10.1111/j.1365-2818.2010.03408.x.